\documentclass[%
 reprint,
superscriptaddress,
nofootinbib,
 amsmath,amssymb,
 aps,
]{revtex4-1}

\usepackage{graphicx}% Include figure files
\usepackage{dcolumn}% Align table columns on decimal point
\usepackage{bm}% bold math
\usepackage{siunitx}
\usepackage{hyperref}% add hypertext capabilities
\usepackage{float}
\usepackage{amsmath}
\usepackage{svg}
\usepackage{newtxmath}

\begin{document}

\title{Few-mode to mesoscopic junctions in gatemon qubits  }

\author{Alisa~Danilenko}
\thanks{These authors contributed equally to this work}
\affiliation{Center for Quantum Devices, Niels Bohr Institute, University of Copenhagen, 2100 Copenhagen, Denmark}

\author{Deividas~Sabonis}
\thanks{These authors contributed equally to this work}

\affiliation{Center for Quantum Devices, Niels Bohr Institute, University of Copenhagen, 2100 Copenhagen, Denmark}

\author{Georg W.~Winkler}
\affiliation{Microsoft Quantum, Microsoft Station Q, University of California, Santa Barbara, California 93106-6105, USA}
 
\author{Oscar~Erlandsson}
\affiliation{Center for Quantum Devices, Niels Bohr Institute, University of Copenhagen, 2100 Copenhagen, Denmark}

\author{Peter~Krogstrup}
\affiliation{Center for Quantum Devices, Niels Bohr Institute, University of Copenhagen, 2100 Copenhagen, Denmark}
\affiliation{Microsoft Quantum Materials Lab--Copenhagen, 2800 Lyngby, Denmark}

\author{Charles~M.~Marcus}
\affiliation{Center for Quantum Devices, Niels Bohr Institute, University of Copenhagen, 2100 Copenhagen, Denmark}

%\date{\today}% 

\begin{abstract}

We investigate a semiconductor nanowire-based gatemon qubit with epitaxial Al on two facets of the nanowire, allowing gate control of wire density. Two segments have the Al removed, one forming a Josephson junction, and the other operating as a transistor, providing {\it in-situ} switching between dc transport and qubit operation. Gating the NW changes the bulk wire potential distribution, while gating the Josephson junction changes the number of junction modes. Both effects are revealed by the dependence of qubit frequency on parallel magnetic field. A detailed model of the wire and junction yields behavior consistent with experiment. In the multi-mode regime, fluctuations in qubit frequency are considerably smaller than the theoretical ``universal" value, also smaller than numerics, and consistent with previous measurements of fluctuating critical current.

\end{abstract}

\maketitle

\begin{figure} 
  \centering
  \includegraphics[width=0.5\textwidth]{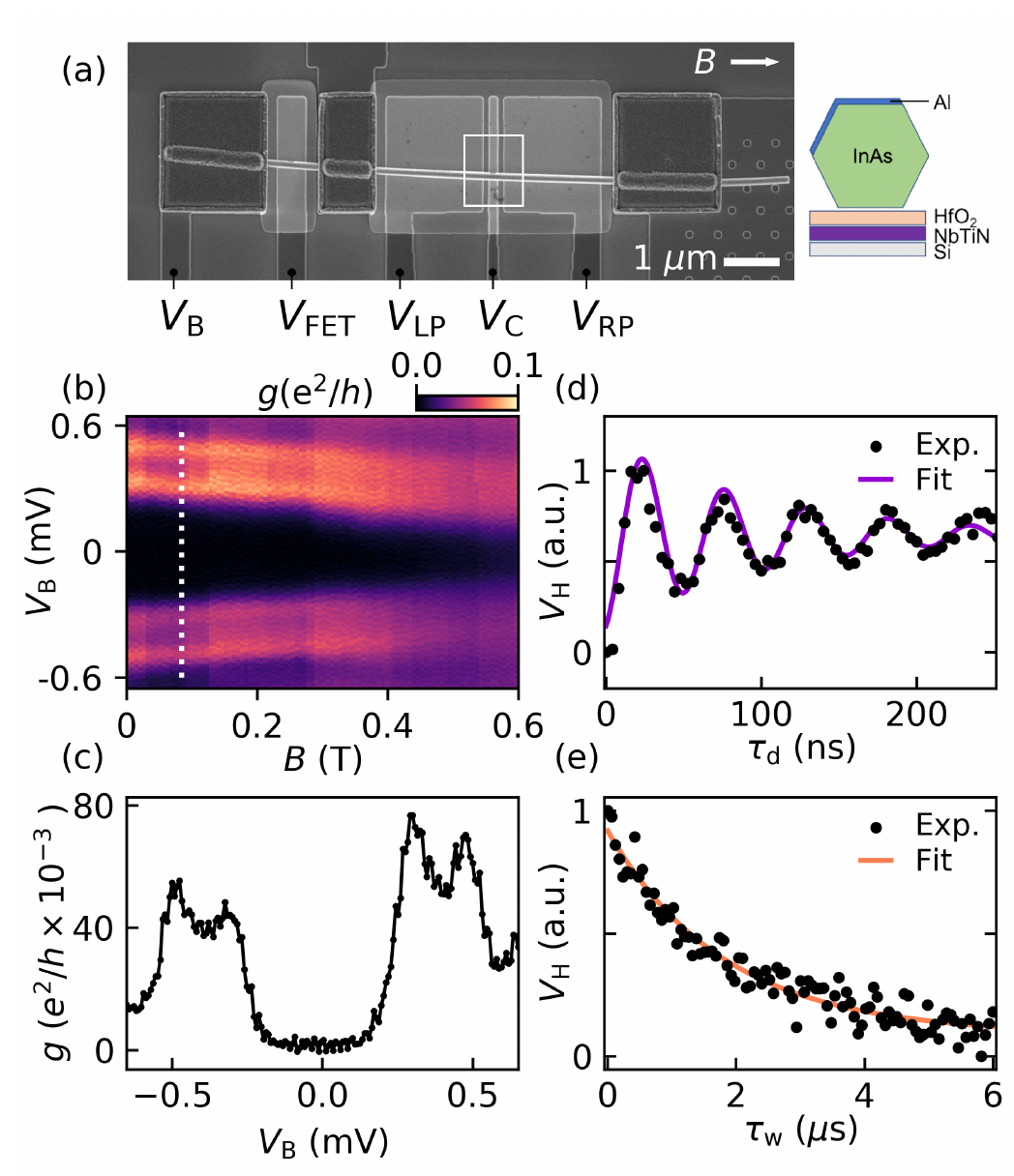}
\caption{(a) Scanning electron micrograph of Device 1 (cross section sketch on the right). The NW is contacted on the right to the ground plane, the capacitor island is connected in the center.
The purpose of gate voltage $V_{\rm  C}$ is to control the Josephson junction, while $V$\textsubscript{LP} and $V$\textsubscript{RP} are intended to tune the bulk wire. Direction of applied magnetic field $B$ is shown. 
(b) Differential conductance $g \equiv \frac{{\mathrm{d}}I_\text{B}}{{\mathrm{d}}V_\text{B}}$ as a function of $V$\textsubscript{B} shows the superconducting gap of the junction in applied field, with a cut (c) taken at $B$~=~0.08~T. $V$\textsubscript{FET}~=~+6~V, $V$\textsubscript{LP}~=~$V$\textsubscript{RP}~=~0~V, $V_{\rm  C}$~=~$-3.6$~V. 
(d)  Rabi oscillations at $B = 0$. Demodulated transmission $V$\textsubscript{H} measured as a function of drive duration $\tau$\textsubscript{d} at the qubit frequency $f_{Q}$~=~4.37~GHz. Exponentially decaying sinusoidal fit  yields Rabi time $T^\text{R}_2~=~120$~ns. (e) Qubit relaxation: $V$\textsubscript{H} measured as a function of wait time $\tau$\textsubscript{w} between drive and readout pulses. The qubit is excited with a $\pi$ pulse calibrated from (d). Exponential fit yields $T_1~=~1.8~\mu$s. 
}
\label{fig:figure1}
\end{figure}

Recent materials advances \cite{KrogstrupEpi} have lead to a new approach to Josephson qubit technology based on hybrid superconductor-semiconductor nanowires (NWs) \cite{de2015realization, Larsen2015} and comparable two-dimensional platforms \cite{Casparis2018}. This approach allows voltage control of qubit operation and reduced sensitivity to charge noise \cite{Kringhoj2020, Kringhoj2020dispersion, Bargerbos2020}. Hybrid NWs can also form the basis of Andreev qubits \cite{hays2018direct, Tosi2019, Hays2021, Canadas2021}, protected $0-\pi$ qubits \cite{Larsen2020}, systems to investigate the presence of topological phases \cite{SabonisLittleParks}, and voltage-controlled qubit couplers \cite{Casparis2019}. Because the electron wavelength in the semiconductor is comparable to the NW diameter, electronic states under the proximitizing superconductor typically occupy a small number of transverse modes \cite{Kringhoj2018}. For NWs with facets not covered by the superconductor, this mode structure can be altered by electrostatic gating \cite{splitthoff22}.

In this Letter,  we compare the magnetic field and gate-voltage dependence of gatemon qubits fabricated from epitaxial InAs/Al NWs to a detailed numerical model of the wire and Andreev bound states (ABSs) in the junction \cite{winkler2019}. To gather parameters for the model, we take advantage of another feature of hybrid NWs by creating a local field-effect transistor (FET) that allows {\it in-situ} switching between dc transport and circuit quantum electrodynamics (cQED) configurations \cite{Kringhoj2020}. Magnetic field and gate-voltage dependences of qubit frequency $f_{Q}$ \cite{AndersHalfshell} are in reasonable agreement with the model, and consistent with gate-voltage \cite{Zou_2017} and magnetic field dependences \cite{guel2014} of critical currents in NW junctions, here measurable in the same qubit junction. 

At gate voltages corresponding to several ABSs in the qubit junction, mesoscopic (random, repeatable) fluctuations of qubit frequency as a function of gate voltage were observed. Comparing experimental results with both numerics and theoretical universal statistics for mesoscopic critical current fluctuations \cite{Beenakkerucf}, we find that the observed qubit-frequency fluctuations, $\sigma_{f_Q}\sim\,$130 MHz, corresponding to critical-current fluctuations $\sigma_{I_c}\sim\,$1~nA, are smaller than theoretical values for a short junction \cite{Beenakkerucf}, though consistent with previous experimental values of critical current fluctuations \cite{Takayanagimeso}. (Mesoscopic fluctuations of $f_Q$ have not been reported previously, to our knowledge). We tentatively ascribed the reduced fluctuation of $f_Q$ to non-ideal material interfaces \cite{dohmeso,Takayanagimeso} or a Fermi velocity mismatch between the Al-covered region and the bare semiconductor junction \cite{shapersmeso} leading to normal reflection competing with Andreev reflection within the junction interfaces. 

Devices were fabricated on a high-resistivity silicon substrate covered with a 20 nm NbTiN film, deposited by sputtering. Each chip contains three gatemons based on NWs , about 100 nm in diameter, with two or three of six facets covered with Al \cite{AndersHalfshell}, two with FET-switched dc transport capabilities, each with individual resonator readout circuits. Overall, three devices were measured. Data from two devices, yielding consistent results, are reported here. Resonators, transmission line, and electrostatic gates were
fabricated using additional layers of sputtered NbTiN, patterned using electron beam lithography and reactive ion etching. Before placing the NW on the bottom gates, a lithographically patterned layer of HfO\textsubscript{2} dielectric was deposited by atomic layer deposition. A micrograph of one of the FET-switched devices is shown in Fig.~\ref{fig:figure1}(a). The right side of the NW connects to the ground plane, the left side to a dc contact through the FET, and the center to the qubit island. 
The Josephson junction, seen inside the white box in Fig.~\ref{fig:figure1}(a), is formed by wet etching $\sim100$~nm of the Al shell. While the orientation of the shell is not discernible during manual NW placement, scanning electron micrographs taken afterwards can resolve the Al shell. Devices with the shell on the up-facing half of the NW, which allow control of carrier density in the NW by bottom gates, are then measured. Gate voltage $V_{\rm  C}$, underneath the junction, was used to tune $f_{Q}$, while voltages $V$\textsubscript{LP} and $V$\textsubscript{RP} were used to tune the density in the bulk NW. All measurements were performed in a dilution refrigerator with a base temperature of 20 mK using a 6-1-1 T vector magnet.

\begin{figure*}
    \centering
    \includegraphics[width=1\textwidth]{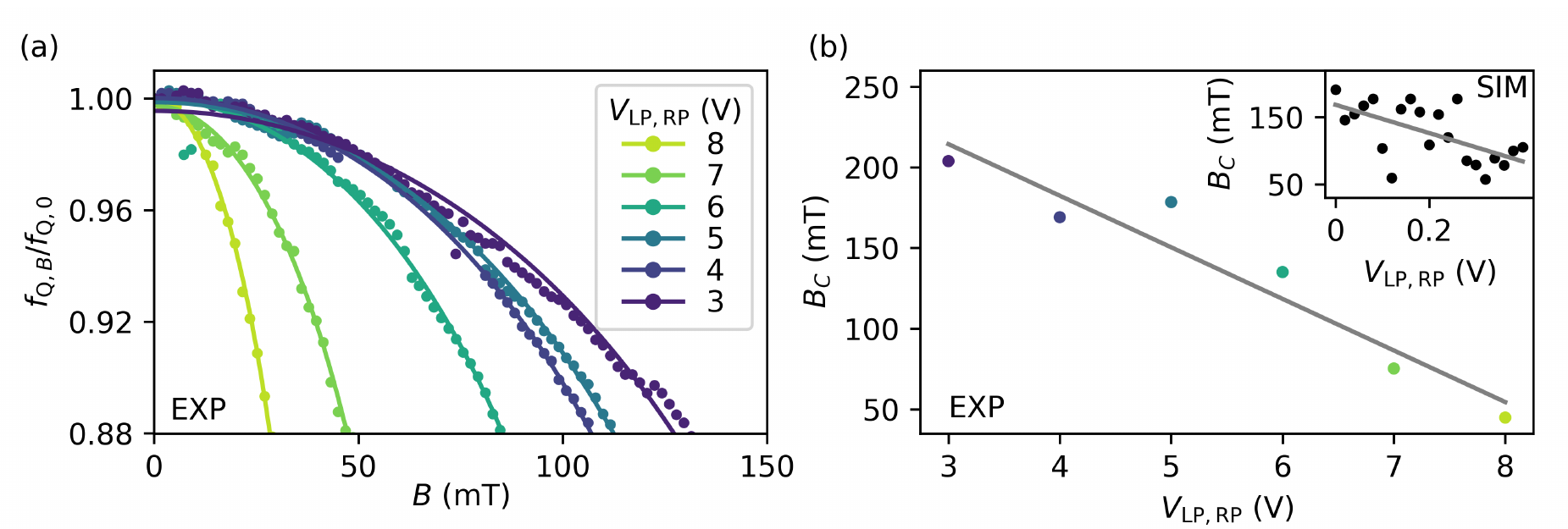}
    \caption{(a) Measured qubit frequency (normalized) as a function of parallel magnetic field $B$ at a range of $V\textsubscript{LP, RP}$ values. $V\textsubscript{C}$ was compensated to keep the zero field frequency constant for all gate configurations. (b) Critical field $B_c$, extracted from fitting the data for each gate configuration measured. Shown with a linear fit.
}
    \label{fig:figure2}
\end{figure*}

Setting the FET in the conducting state by applying +6.0 V on the FET gate, differential conductance $g \equiv {\mathrm{d}}I_\text{B}/{\mathrm{d}}V_\text{B}$ of the junction was measured as a function of voltage bias $ V_\text{B}$. With the qubit junction in the tunneling regime, $g$ can be used to measure the parent gap and ABS features in the semiconductor, as shown in Figs.~\ref{fig:figure1}(b, c). The two higher-bias peaks occur at the bias where coherence peaks from the Al gap, $\Delta\sim 250\, \mu$eV, on the two sides of the junctions align, $V_{\text B}\sim\pm 2\Delta/e\sim \pm 0.5\,$mV. 
The two lower-bias peaks reflect where the coherence peak in one lead aligns with a subgap ABS in the other lead.  Extrapolating the field dependence of the gap gives a field of $\sim$1.4 T where the gap closes.
Following dc transport characterization of the junction, the FET was switched to a non-conductive state (FET gate at $-6$ V) allowing the device to be operated in cQED mode as a qubit. Setting $V\textsubscript{C} = -1.7$~V, $V\textsubscript{LP, RP} = 0$~V gives a qubit frequency $f_{Q} = 4.37$~GHz, measured via two-tone spectroscopy. Rabi oscillations [Fig. \ref{fig:figure1}(d)] were measured by applying a series of pulses of duration $\tau\textsubscript{d}$ at $f_{Q}$ and plotting the demodulated transmission $V\textsubscript{H}$ as a function of $\tau\textsubscript{d}$. Fitting to an exponentially decaying sinusoid yields a Rabi time of 119~$\pm$~1~ns. Qubit relaxation [Fig.~\ref{fig:figure1}(e)] was measured by applying a $\pi$ pulse, found using data in (d), at $f_{Q}$, then waiting $\tau\textsubscript{w}$ before applying a readout pulse at the resonator frequency, 5.46~GHz, giving a $V$\textsubscript{H} signal that decreases with increasing $\tau\textsubscript{w}$. An exponential fit yields a qubit lifetime of 1.81~$\pm$~0.13~$\mu$s. \\
The dependence of $f_{Q}$ on axial magnetic field $B$ and gate voltage is shown in Fig.~\ref{fig:figure2}(a), for Device 2, which is similar to Device 1 in material and fabrication. For each value of $V\textsubscript{LP, RP}$, $V\textsubscript{C}$ was compensated to keep the zero-field frequency constant, to try to minimise the effect of $V\textsubscript{LP, RP}$ on the junction, since the aim was to tune the bulk wire density. This compensation had the added benefit of making sure that $f_{Q}$ stayed in a measurable range at zero field. The changes made to $V\textsubscript{C}$ are very small compared to its full range of operation. In the measured field and frequency range, the qubit frequency decreased monotonically with increasing $B$.
The solid lines in Fig.~\ref{fig:figure2}(a) are simple fits relating the qubit frequency expected closing of the superconducting gap in field, $f_{Q}(B)  = f_{Q}(0)\,[{1-(B/B_c)^2}]^{1/4}$, where $B_c$ is the critical field \cite{tinkham2004introduction}. As illustrated in Fig.~\ref{fig:figure2}(b), a trend in the dependence on $V\textsubscript{LP, RP}$ was observed: the more positive the gate voltage applied to the NW, the more rapidly $f_{Q}$ decayed in field.

To understand the effect of the gate voltage configuration on the magnetic field dependence of qubit frequency, we perform numerical modeling of the energy spectrum and qubit frequency using codes similar to Refs.~\cite{SoleScience2020, AndersPRL2021, shen2021parity} using a self-consistent Thomas-Fermi approximation, including Zeeman and orbital effects of the magnetic field, treating coupling of the superconductor to the semiconductor in terms of a self-energy boundary condition (see Supplementary Material).  We find that the simulated qubit frequency follows roughly the quadratic decay of the parent gap for small fields, as observed in the experiment. Figure~\ref{fig:figure2}(b) inset shows the simulated dependence of $B_{C}$ on the NW gate voltage. It should be noted that the lever arm in the simulation is much larger than in experiment, hence the significant difference in absolute gate voltage values. There is still a trend towards more rapidly decaying $f_{Q}$ at more positive gate voltage in the simulation, but the numerical results are less monotonic than what we observe in the experiment. 
 At more negative gate voltages, the wave function is pressed up away from the gates and close to the superconducting shell, reducing the wave function cross-section threaded by magnetic flux, leading to a more gradual decay of $f_{Q}$ in field. Within this interpretation, the effect is due to gating of the NW, not the junction.  

\begin{figure}
\includegraphics[width=0.5\textwidth]{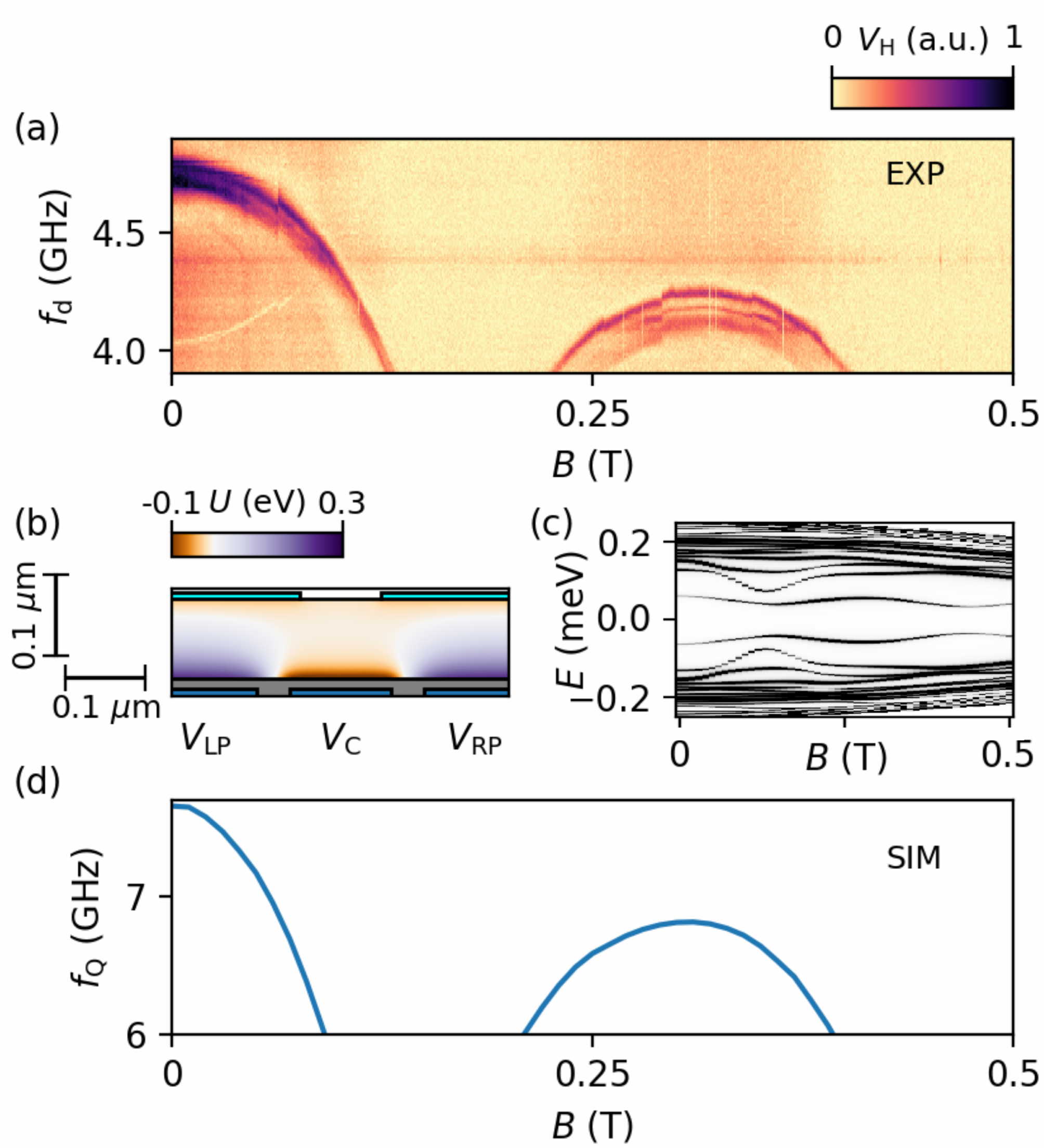}
\caption{Two-tone spectroscopy
as a function of drive frequency, $f$\textsubscript{d}, and parallel magnetic
field, $B$, with  $V$\textsubscript{FET}~=~$+6.0$~V,  $V$\textsubscript{LP, RP}~=~0~V, $V$\textsubscript{C}~=~$-1.8$~V. At low field, the qubit frequency, $f_{Q}$, decreases rapidly, disappearing from the measurement window at $B\sim 130$ mT. At higher field, $f_{Q}$ recovers, re-entering the measurement window at $B \sim 220$ mT then decreases, leaving the measurement window at $B\sim 400$~mT.
(b) Numerical electrostatic potential, $U$, near the junction on a vertical cut through the NW, for a configuration with nonmonotonic $f_{Q}$. The superconductor (Al) is indicated on the top (cyan), due to the positive band offset of 50~meV there is an accumulation layer towards it. The left plunger, the cutter and the right plunger gates are indicated on the bottom (dark blue). Both plungers are set to the same voltage $V_\mathrm{LP}=V_\mathrm{RP}=0\,\mathrm{V}$. $V_\mathrm{C}=+0.55\,\mathrm{V}$. (c) Simulated local density of states in the junction region for this gate configuration as a function of parallel magnetic field. A few low-energy ABSs showing flux-modulated oscillations can be found inside the junction. (d) Simulated qubit frequency for the same gate configuration. The flux-modulation of ABS results in an oscillation of the qubit frequency.}
\label{fig:figure3}
\end{figure}

With increasing magnetic field, nonmonotonic evolution of $f_{Q}$ was observed both in experiment, as shown in Fig.~\ref{fig:figure3}. The corresponding electrostatics of the wire and junction, as well as the field dependence of ABSs in the junction, are shown in  Figs.~\ref{fig:figure3}(b,c). In this configuration, the junction was fairly open, with $V\textsubscript{C}$ accumulating electrons, while the back gates $V\textsubscript{LP, RP}$ were rather negative, pushing electrons towards the superconductor.

The spectrum in Fig.~\ref{fig:figure3}(c) shows several ABS in the junction, a few of which show an oscillatory behavior in field. The magnetic field where they have their minimum corresponds roughly to half a magnetic flux quantum through the cross section of the NW. 

The energies and phase dependence of ABS is directly linked to the qubit frequency. To understand the effect of ABS energies on the qubit frequency, we simulate the supercurrent using the Kwant package~\cite{kwant} and the analysis developed in Refs.~\cite{Ostroukh_2016, Zou_2017}.
The simulated critical current $I$\textsubscript{c}, yields a qubit frequency $f_{\text{Q}}\approx\sqrt{8 E_{J} E_{C}}/h=\sqrt{2E_{C}I_{c}/h\pi e}$ (valid for $E_{J} \gg E_{C}$; in the experiment $E_J/E_C\sim 10$) , where $E_{C}= 2\,\mu$eV is estimated from electrostatic simulations of the qubit island~\cite{comsol} and $E_{J}=\hbar I_{c}/2e$. 

In the simulated qubit frequency as a function of magnetic field [Fig.~\ref{fig:figure3}(d)], one can see a lobe-like structure where the qubit frequency oscillates as a function of magnetic field. The oscillation period of the ABS spectrum is lined up with the qubit frequency. Therefore, we conclude that the oscillatory behavior of the qubit frequency in field is linked to flux modulation of ABSs which form in the junction. Based on these simulations, we propose that the additional revival observed in the experiment is due to junction-based physics, in contrast to the low field behavior of $f_{Q}$ in Fig.~\ref{fig:figure2}, which we attribute to depletion within the wire.

\begin{figure}
\includegraphics[width=0.5\textwidth]{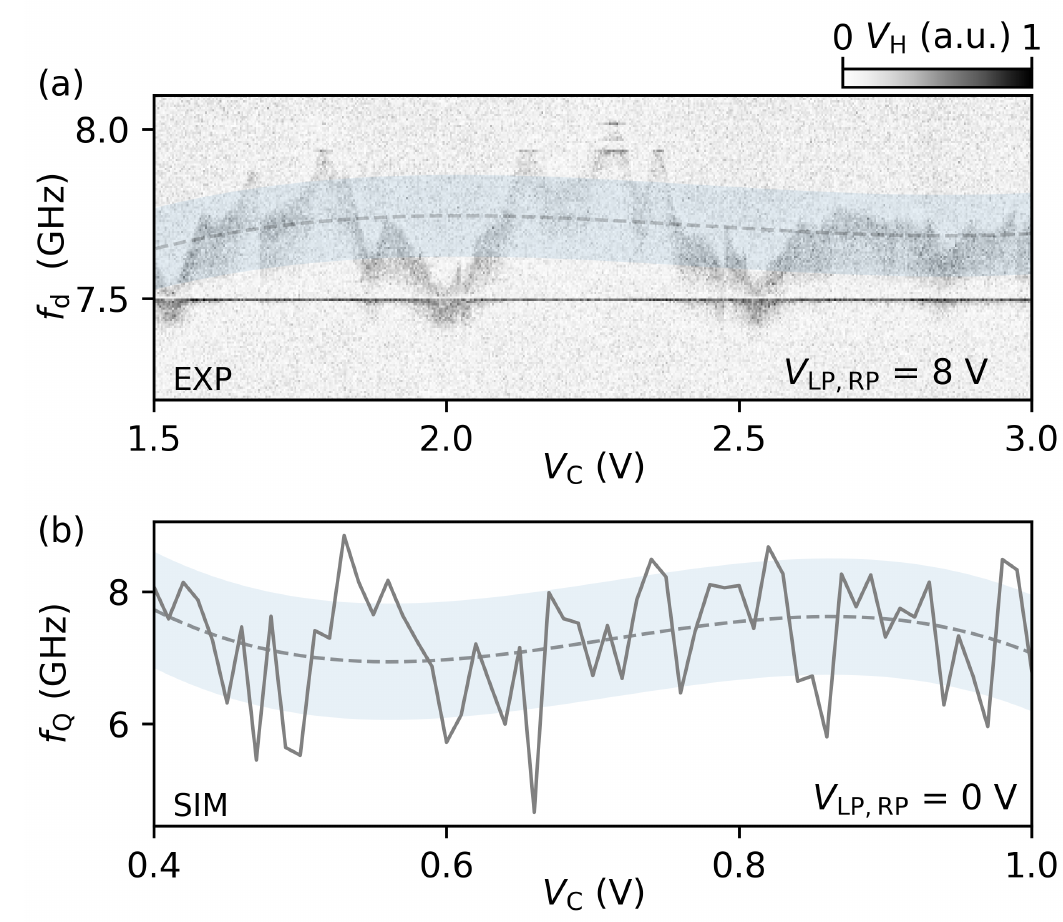}
\caption{(a) Two-tone spectroscopy as a function of drive frequency ($f_{\rm d}$) and gate voltage $V_{\rm  C}$. The qubit frequency exhibits mesoscopic fluctuations. The data is fitted with a smoothed spline (dashed line), and the light blue shading indicates the standard deviation, covering the area within $+/- \sigma$ of the fit. (b) Simulated trace of qubit frequency. $f_{Q}$ as a function of gate voltage $V_{\rm C}$, with the fluctuations analyzed in a similar manner to the experimental data. }
\label{fig:figure4}
\end{figure}

Increasing the number of modes in the semiconductor as much as possible by setting gate voltages $V_{\rm PL}$ and $V_{\rm PR}$ to +8~V results in random-looking but repeatable mesoscopic fluctuations of $f_{Q}$ with $V_{\text C}$. In this regime, $\frac{E_J}{E_C}$ is closer to $\sim~30$. These fluctuations are closely related to mesoscopic fluctuations of critical current in Josephson junctions \cite{Beenakkerucf, Takayanagimeso, gunelmeso}. In superconductor-normal-superconductor (SNS) Josephson junctions with a distance $L$ between superconductors which is long compared to the mean free path, $l$, in the N region and the superconducting coherence length, $\xi$, mesoscopic fluctuations of critical current $I_{\text c}$ are expected to have a non-universal magnitude, with standard deviation  $\sigma_{I_{\text c}}\sim ev_Fl/L^2$, where $v_{F}$ is the Fermi energy in the N region \cite{altspivmeso}.  On the other hand, in short, disordered junctions $l\ll L<\xi$, critical current fluctuations are are expected to be ``universal," $\sigma_{I_{\text c}}\sim e\Delta_{0}$/$\hbar$, independent of  junction parameters \cite{Beenakkerucf}. Our junction length is $\sim$ 100 nm, which we can expect to be in the short junction limit, consistent with previous work \cite{vanWoerkom2016, Spanton2017}.

Fluctuation statistics were extracted from experimental data by sampling over $V_{\text C}$ in Device 2. Figure~\ref{fig:figure4}(a) shows two-tone spectroscopy data as a function of junction gate $V_{\rm  C}$ at $B=0$. We perform these measurements at much more positive gate voltages than in the previous section, such that the number of modes in the junction is maximized. The qubit frequency fluctuation is quantified by fitting a smoothed spline to the trace and extracting a standard deviation, yielding $\sigma_{f_Q}\sim130$~MHz. This corresponds to critical current fluctuation $\sigma_{I_c} \sim (\pi e h f_Q/E_C)\sigma_{f_Q} \sim$ 1.0 nA. This is much smaller than the theoretically predicted value \cite{Beenakkerucf} for a short junction, which would give of order $\sim$ 50 nA, but is in closer agreement with experiments measuring critical current oscillations \cite{Takayanagimeso, gunelmeso} which find fluctuations on the order of 1 nA.

Similar traces are simulated for a range of gate voltages, as shown in Fig.~\ref{fig:figure4}(b) for simulated $V_{\rm{LP, RP}} = 0$. Changing the plunger gate voltages can be viewed as a variation in the simulated properties of the junction, which was not attempted in our experiment. The simulated fluctuation in $f_Q$ are in the range 0.4--1~GHz, corresponding to critical current fluctuations in the range  3--9~nA. These values are larger than what was observed experimentally, but smaller than the universal value for a short junction. \\\\

In conclusion, we have investigated semiconductor based transmon qubits that allow both dc transport and cQED operation, controlled by a field effect transistor, and are magnetic field compatible. With the help of numerics, we observe that the rate of qubit frequency decay in field is controllable by gating of the bulk NW, while additional oscillatory behavior at higher fields may be attributed to flux modulation of ABSs in the junction. 
In the many-channel regime, mesoscopic fluctuations of the qubit frequency were considerably smaller than expected from universal theory, somewhat smaller than numerics, and comparable to previous corresponding results of fluctuations in critical current. 

\section*{Acknowledgments}

We thank Ruben Grigoryan, Bernard van Heck, Marina Hesselberg, Karthik Jambunathan, Anders Kringh\o j, Thorvald W. Larsen, Robert McNeil, Karolis Parfeniukas,  Karl Petersson, Lukas Splitthoff, Agnieszka Telecka, Shivendra Upadhyay, Sachin Yadav, and David van Zanten for help with device fabrication and valuable discussion. Research was supported by the Danish National Research Foundation, Microsoft, a grant (Project 43951) from VILLUM FONDEN, and the European Research Council under grant HEMs-DAM No.716655.

\bibliography{refs}

%merlin.mbs apsrev4-1.bst 2010-07-25 4.21a (PWD, AO, DPC) hacked
%Control: key (0)
%Control: author (8) initials jnrlst
%Control: editor formatted (1) identically to author
%Control: production of article title (-1) disabled
%Control: page (0) single
%Control: year (1) truncated
%Control: production of eprint (0) enabled
\begin{thebibliography}{35}%
\makeatletter
\providecommand \@ifxundefined [1]{%
 \@ifx{#1\undefined}
}%
\providecommand \@ifnum [1]{%
 \ifnum #1\expandafter \@firstoftwo
 \else \expandafter \@secondoftwo
 \fi
}%
\providecommand \@ifx [1]{%
 \ifx #1\expandafter \@firstoftwo
 \else \expandafter \@secondoftwo
 \fi
}%
\providecommand \natexlab [1]{#1}%
\providecommand \enquote  [1]{``#1''}%
\providecommand \bibnamefont  [1]{#1}%
\providecommand \bibfnamefont [1]{#1}%
\providecommand \citenamefont [1]{#1}%
\providecommand \href@noop [0]{\@secondoftwo}%
\providecommand \href [0]{\begingroup \@sanitize@url \@href}%
\providecommand \@href[1]{\@@startlink{#1}\@@href}%
\providecommand \@@href[1]{\endgroup#1\@@endlink}%
\providecommand \@sanitize@url [0]{\catcode `\\12\catcode `\$12\catcode
  `\&12\catcode `\#12\catcode `\^12\catcode `\_12\catcode `\%12\relax}%
\providecommand \@@startlink[1]{}%
\providecommand \@@endlink[0]{}%
\providecommand \url  [0]{\begingroup\@sanitize@url \@url }%
\providecommand \@url [1]{\endgroup\@href {#1}{\urlprefix }}%
\providecommand \urlprefix  [0]{URL }%
\providecommand \Eprint [0]{\href }%
\providecommand \doibase [0]{http://dx.doi.org/}%
\providecommand \selectlanguage [0]{\@gobble}%
\providecommand \bibinfo  [0]{\@secondoftwo}%
\providecommand \bibfield  [0]{\@secondoftwo}%
\providecommand \translation [1]{[#1]}%
\providecommand \BibitemOpen [0]{}%
\providecommand \bibitemStop [0]{}%
\providecommand \bibitemNoStop [0]{.\EOS\space}%
\providecommand \EOS [0]{\spacefactor3000\relax}%
\providecommand \BibitemShut  [1]{\csname bibitem#1\endcsname}%
\let\auto@bib@innerbib\@empty
%</preamble>
\bibitem [{\citenamefont {Krogstrup}\ \emph {et~al.}(2015)\citenamefont
  {Krogstrup}, \citenamefont {Ziino}, \citenamefont {Chang}, \citenamefont
  {Albrecht}, \citenamefont {Madsen}, \citenamefont {Johnson}, \citenamefont
  {Nyg{\aa}rd}, \citenamefont {Marcus},\ and\ \citenamefont
  {Jespersen}}]{KrogstrupEpi}%
  \BibitemOpen
  \bibfield  {author} {\bibinfo {author} {\bibfnamefont {P.}~\bibnamefont
  {Krogstrup}}, \bibinfo {author} {\bibfnamefont {N.}~\bibnamefont {Ziino}},
  \bibinfo {author} {\bibfnamefont {W.}~\bibnamefont {Chang}}, \bibinfo
  {author} {\bibfnamefont {S.}~\bibnamefont {Albrecht}}, \bibinfo {author}
  {\bibfnamefont {M.}~\bibnamefont {Madsen}}, \bibinfo {author} {\bibfnamefont
  {E.}~\bibnamefont {Johnson}}, \bibinfo {author} {\bibfnamefont
  {J.}~\bibnamefont {Nyg{\aa}rd}}, \bibinfo {author} {\bibfnamefont
  {C.}~\bibnamefont {Marcus}}, \ and\ \bibinfo {author} {\bibfnamefont
  {T.}~\bibnamefont {Jespersen}},\ }\href {\doibase 10.1038/nmat4176}
  {\bibfield  {journal} {\bibinfo  {journal} {Nature Materials}\ }\textbf
  {\bibinfo {volume} {14}},\ \bibinfo {pages} {400} (\bibinfo {year}
  {2015})}\BibitemShut {NoStop}%
\bibitem [{\citenamefont {de~Lange}\ \emph {et~al.}(2015)\citenamefont
  {de~Lange}, \citenamefont {van Heck}, \citenamefont {Bruno}, \citenamefont
  {van Woerkom}, \citenamefont {Geresdi}, \citenamefont {Plissard},
  \citenamefont {Bakkers}, \citenamefont {Akhmerov},\ and\ \citenamefont
  {DiCarlo}}]{de2015realization}%
  \BibitemOpen
  \bibfield  {author} {\bibinfo {author} {\bibfnamefont {G.}~\bibnamefont
  {de~Lange}}, \bibinfo {author} {\bibfnamefont {B.}~\bibnamefont {van Heck}},
  \bibinfo {author} {\bibfnamefont {A.}~\bibnamefont {Bruno}}, \bibinfo
  {author} {\bibfnamefont {D.~J.}\ \bibnamefont {van Woerkom}}, \bibinfo
  {author} {\bibfnamefont {A.}~\bibnamefont {Geresdi}}, \bibinfo {author}
  {\bibfnamefont {S.~R.}\ \bibnamefont {Plissard}}, \bibinfo {author}
  {\bibfnamefont {E.~P. A.~M.}\ \bibnamefont {Bakkers}}, \bibinfo {author}
  {\bibfnamefont {A.~R.}\ \bibnamefont {Akhmerov}}, \ and\ \bibinfo {author}
  {\bibfnamefont {L.}~\bibnamefont {DiCarlo}},\ }\href {\doibase
  10.1103/PhysRevLett.115.127002} {\bibfield  {journal} {\bibinfo  {journal}
  {Phys. Rev. Lett.}\ }\textbf {\bibinfo {volume} {115}},\ \bibinfo {pages}
  {127002} (\bibinfo {year} {2015})}\BibitemShut {NoStop}%
\bibitem [{\citenamefont {Larsen}\ \emph {et~al.}(2015)\citenamefont {Larsen},
  \citenamefont {Petersson}, \citenamefont {Kuemmeth}, \citenamefont
  {Jespersen}, \citenamefont {Krogstrup}, \citenamefont {Nyg\aa{}rd},\ and\
  \citenamefont {Marcus}}]{Larsen2015}%
  \BibitemOpen
  \bibfield  {author} {\bibinfo {author} {\bibfnamefont {T.~W.}\ \bibnamefont
  {Larsen}}, \bibinfo {author} {\bibfnamefont {K.~D.}\ \bibnamefont
  {Petersson}}, \bibinfo {author} {\bibfnamefont {F.}~\bibnamefont {Kuemmeth}},
  \bibinfo {author} {\bibfnamefont {T.~S.}\ \bibnamefont {Jespersen}}, \bibinfo
  {author} {\bibfnamefont {P.}~\bibnamefont {Krogstrup}}, \bibinfo {author}
  {\bibfnamefont {J.}~\bibnamefont {Nyg\aa{}rd}}, \ and\ \bibinfo {author}
  {\bibfnamefont {C.~M.}\ \bibnamefont {Marcus}},\ }\href {\doibase
  10.1103/PhysRevLett.115.127001} {\bibfield  {journal} {\bibinfo  {journal}
  {Phys. Rev. Lett.}\ }\textbf {\bibinfo {volume} {115}},\ \bibinfo {pages}
  {127001} (\bibinfo {year} {2015})}\BibitemShut {NoStop}%
\bibitem [{\citenamefont {Casparis}\ \emph {et~al.}(2018)\citenamefont
  {Casparis}, \citenamefont {Connolly}, \citenamefont {Kjaergaard},
  \citenamefont {Pearson}, \citenamefont {Kringhøj}, \citenamefont {Larsen},
  \citenamefont {Kuemmeth}, \citenamefont {Wang}, \citenamefont {Thomas},
  \citenamefont {Gronin}, \citenamefont {Gardner}, \citenamefont {Manfra},
  \citenamefont {Marcus},\ and\ \citenamefont {Petersson}}]{Casparis2018}%
  \BibitemOpen
  \bibfield  {author} {\bibinfo {author} {\bibfnamefont {L.}~\bibnamefont
  {Casparis}}, \bibinfo {author} {\bibfnamefont {M.~R.}\ \bibnamefont
  {Connolly}}, \bibinfo {author} {\bibfnamefont {M.}~\bibnamefont
  {Kjaergaard}}, \bibinfo {author} {\bibfnamefont {N.~J.}\ \bibnamefont
  {Pearson}}, \bibinfo {author} {\bibfnamefont {A.}~\bibnamefont {Kringhøj}},
  \bibinfo {author} {\bibfnamefont {T.~W.}\ \bibnamefont {Larsen}}, \bibinfo
  {author} {\bibfnamefont {F.}~\bibnamefont {Kuemmeth}}, \bibinfo {author}
  {\bibfnamefont {T.}~\bibnamefont {Wang}}, \bibinfo {author} {\bibfnamefont
  {C.}~\bibnamefont {Thomas}}, \bibinfo {author} {\bibfnamefont
  {S.}~\bibnamefont {Gronin}}, \bibinfo {author} {\bibfnamefont {G.~C.}\
  \bibnamefont {Gardner}}, \bibinfo {author} {\bibfnamefont {M.~J.}\
  \bibnamefont {Manfra}}, \bibinfo {author} {\bibfnamefont {C.~M.}\
  \bibnamefont {Marcus}}, \ and\ \bibinfo {author} {\bibfnamefont {K.~D.}\
  \bibnamefont {Petersson}},\ }\href {\doibase 10.1038/s41565-018-0207-y}
  {\bibfield  {journal} {\bibinfo  {journal} {Nature Nanotechnology}\ }\textbf
  {\bibinfo {volume} {13}},\ \bibinfo {pages} {915} (\bibinfo {year} {2018})},\
  \Eprint {http://arxiv.org/abs/1711.07665} {1711.07665} \BibitemShut {NoStop}%
\bibitem [{\citenamefont {Kringh\o{}j}\ \emph
  {et~al.}(2020{\natexlab{a}})\citenamefont {Kringh\o{}j}, \citenamefont
  {Larsen}, \citenamefont {van Heck}, \citenamefont {Sabonis}, \citenamefont
  {Erlandsson}, \citenamefont {Petkovic}, \citenamefont {Pikulin},
  \citenamefont {Krogstrup}, \citenamefont {Petersson},\ and\ \citenamefont
  {Marcus}}]{Kringhoj2020}%
  \BibitemOpen
  \bibfield  {author} {\bibinfo {author} {\bibfnamefont {A.}~\bibnamefont
  {Kringh\o{}j}}, \bibinfo {author} {\bibfnamefont {T.~W.}\ \bibnamefont
  {Larsen}}, \bibinfo {author} {\bibfnamefont {B.}~\bibnamefont {van Heck}},
  \bibinfo {author} {\bibfnamefont {D.}~\bibnamefont {Sabonis}}, \bibinfo
  {author} {\bibfnamefont {O.}~\bibnamefont {Erlandsson}}, \bibinfo {author}
  {\bibfnamefont {I.}~\bibnamefont {Petkovic}}, \bibinfo {author}
  {\bibfnamefont {D.~I.}\ \bibnamefont {Pikulin}}, \bibinfo {author}
  {\bibfnamefont {P.}~\bibnamefont {Krogstrup}}, \bibinfo {author}
  {\bibfnamefont {K.~D.}\ \bibnamefont {Petersson}}, \ and\ \bibinfo {author}
  {\bibfnamefont {C.~M.}\ \bibnamefont {Marcus}},\ }\href {\doibase
  10.1103/PhysRevLett.124.056801} {\bibfield  {journal} {\bibinfo  {journal}
  {Phys. Rev. Lett.}\ }\textbf {\bibinfo {volume} {124}},\ \bibinfo {pages}
  {056801} (\bibinfo {year} {2020}{\natexlab{a}})}\BibitemShut {NoStop}%
\bibitem [{\citenamefont {Kringh\o{}j}\ \emph
  {et~al.}(2020{\natexlab{b}})\citenamefont {Kringh\o{}j}, \citenamefont {van
  Heck}, \citenamefont {Larsen}, \citenamefont {Erlandsson}, \citenamefont
  {Sabonis}, \citenamefont {Krogstrup}, \citenamefont {Casparis}, \citenamefont
  {Petersson},\ and\ \citenamefont {Marcus}}]{Kringhoj2020dispersion}%
  \BibitemOpen
  \bibfield  {author} {\bibinfo {author} {\bibfnamefont {A.}~\bibnamefont
  {Kringh\o{}j}}, \bibinfo {author} {\bibfnamefont {B.}~\bibnamefont {van
  Heck}}, \bibinfo {author} {\bibfnamefont {T.~W.}\ \bibnamefont {Larsen}},
  \bibinfo {author} {\bibfnamefont {O.}~\bibnamefont {Erlandsson}}, \bibinfo
  {author} {\bibfnamefont {D.}~\bibnamefont {Sabonis}}, \bibinfo {author}
  {\bibfnamefont {P.}~\bibnamefont {Krogstrup}}, \bibinfo {author}
  {\bibfnamefont {L.}~\bibnamefont {Casparis}}, \bibinfo {author}
  {\bibfnamefont {K.~D.}\ \bibnamefont {Petersson}}, \ and\ \bibinfo {author}
  {\bibfnamefont {C.~M.}\ \bibnamefont {Marcus}},\ }\href {\doibase
  10.1103/PhysRevLett.124.246803} {\bibfield  {journal} {\bibinfo  {journal}
  {Phys. Rev. Lett.}\ }\textbf {\bibinfo {volume} {124}},\ \bibinfo {pages}
  {246803} (\bibinfo {year} {2020}{\natexlab{b}})}\BibitemShut {NoStop}%
\bibitem [{\citenamefont {Bargerbos}\ \emph {et~al.}(2020)\citenamefont
  {Bargerbos}, \citenamefont {Uilhoorn}, \citenamefont {Yang}, \citenamefont
  {Krogstrup}, \citenamefont {Kouwenhoven}, \citenamefont {de~Lange},
  \citenamefont {van Heck},\ and\ \citenamefont {Kou}}]{Bargerbos2020}%
  \BibitemOpen
  \bibfield  {author} {\bibinfo {author} {\bibfnamefont {A.}~\bibnamefont
  {Bargerbos}}, \bibinfo {author} {\bibfnamefont {W.}~\bibnamefont {Uilhoorn}},
  \bibinfo {author} {\bibfnamefont {C.-K.}\ \bibnamefont {Yang}}, \bibinfo
  {author} {\bibfnamefont {P.}~\bibnamefont {Krogstrup}}, \bibinfo {author}
  {\bibfnamefont {L.~P.}\ \bibnamefont {Kouwenhoven}}, \bibinfo {author}
  {\bibfnamefont {G.}~\bibnamefont {de~Lange}}, \bibinfo {author}
  {\bibfnamefont {B.}~\bibnamefont {van Heck}}, \ and\ \bibinfo {author}
  {\bibfnamefont {A.}~\bibnamefont {Kou}},\ }\href {\doibase
  10.1103/PhysRevLett.124.246802} {\bibfield  {journal} {\bibinfo  {journal}
  {Phys. Rev. Lett.}\ }\textbf {\bibinfo {volume} {124}},\ \bibinfo {pages}
  {246802} (\bibinfo {year} {2020})}\BibitemShut {NoStop}%
\bibitem [{\citenamefont {Hays}\ \emph {et~al.}(2018)\citenamefont {Hays},
  \citenamefont {de~Lange}, \citenamefont {Serniak}, \citenamefont {van
  Woerkom}, \citenamefont {Bouman}, \citenamefont {Krogstrup}, \citenamefont
  {Nyg\aa{}rd}, \citenamefont {Geresdi},\ and\ \citenamefont
  {Devoret}}]{hays2018direct}%
  \BibitemOpen
  \bibfield  {author} {\bibinfo {author} {\bibfnamefont {M.}~\bibnamefont
  {Hays}}, \bibinfo {author} {\bibfnamefont {G.}~\bibnamefont {de~Lange}},
  \bibinfo {author} {\bibfnamefont {K.}~\bibnamefont {Serniak}}, \bibinfo
  {author} {\bibfnamefont {D.~J.}\ \bibnamefont {van Woerkom}}, \bibinfo
  {author} {\bibfnamefont {D.}~\bibnamefont {Bouman}}, \bibinfo {author}
  {\bibfnamefont {P.}~\bibnamefont {Krogstrup}}, \bibinfo {author}
  {\bibfnamefont {J.}~\bibnamefont {Nyg\aa{}rd}}, \bibinfo {author}
  {\bibfnamefont {A.}~\bibnamefont {Geresdi}}, \ and\ \bibinfo {author}
  {\bibfnamefont {M.~H.}\ \bibnamefont {Devoret}},\ }\href {\doibase
  10.1103/PhysRevLett.121.047001} {\bibfield  {journal} {\bibinfo  {journal}
  {Phys. Rev. Lett.}\ }\textbf {\bibinfo {volume} {121}},\ \bibinfo {pages}
  {047001} (\bibinfo {year} {2018})}\BibitemShut {NoStop}%
\bibitem [{\citenamefont {Tosi}\ \emph {et~al.}(2019)\citenamefont {Tosi},
  \citenamefont {Metzger}, \citenamefont {Goffman}, \citenamefont {Urbina},
  \citenamefont {Pothier}, \citenamefont {Park}, \citenamefont {Yeyati},
  \citenamefont {Nyg\aa{}rd},\ and\ \citenamefont {Krogstrup}}]{Tosi2019}%
  \BibitemOpen
  \bibfield  {author} {\bibinfo {author} {\bibfnamefont {L.}~\bibnamefont
  {Tosi}}, \bibinfo {author} {\bibfnamefont {C.}~\bibnamefont {Metzger}},
  \bibinfo {author} {\bibfnamefont {M.~F.}\ \bibnamefont {Goffman}}, \bibinfo
  {author} {\bibfnamefont {C.}~\bibnamefont {Urbina}}, \bibinfo {author}
  {\bibfnamefont {H.}~\bibnamefont {Pothier}}, \bibinfo {author} {\bibfnamefont
  {S.}~\bibnamefont {Park}}, \bibinfo {author} {\bibfnamefont {A.~L.}\
  \bibnamefont {Yeyati}}, \bibinfo {author} {\bibfnamefont {J.}~\bibnamefont
  {Nyg\aa{}rd}}, \ and\ \bibinfo {author} {\bibfnamefont {P.}~\bibnamefont
  {Krogstrup}},\ }\href {\doibase 10.1103/PhysRevX.9.011010} {\bibfield
  {journal} {\bibinfo  {journal} {Phys. Rev. X}\ }\textbf {\bibinfo {volume}
  {9}},\ \bibinfo {pages} {011010} (\bibinfo {year} {2019})}\BibitemShut
  {NoStop}%
\bibitem [{\citenamefont {Hays}\ \emph {et~al.}(2021)\citenamefont {Hays},
  \citenamefont {Fatemi}, \citenamefont {Bouman}, \citenamefont {Cerrillo},
  \citenamefont {Diamond}, \citenamefont {Serniak}, \citenamefont {Connolly},
  \citenamefont {Krogstrup}, \citenamefont {Nygård}, \citenamefont {Yeyati},
  \citenamefont {Geresdi},\ and\ \citenamefont {Devoret}}]{Hays2021}%
  \BibitemOpen
  \bibfield  {author} {\bibinfo {author} {\bibfnamefont {M.}~\bibnamefont
  {Hays}}, \bibinfo {author} {\bibfnamefont {V.}~\bibnamefont {Fatemi}},
  \bibinfo {author} {\bibfnamefont {D.}~\bibnamefont {Bouman}}, \bibinfo
  {author} {\bibfnamefont {J.}~\bibnamefont {Cerrillo}}, \bibinfo {author}
  {\bibfnamefont {S.}~\bibnamefont {Diamond}}, \bibinfo {author} {\bibfnamefont
  {K.}~\bibnamefont {Serniak}}, \bibinfo {author} {\bibfnamefont
  {T.}~\bibnamefont {Connolly}}, \bibinfo {author} {\bibfnamefont
  {P.}~\bibnamefont {Krogstrup}}, \bibinfo {author} {\bibfnamefont
  {J.}~\bibnamefont {Nygård}}, \bibinfo {author} {\bibfnamefont {A.~L.}\
  \bibnamefont {Yeyati}}, \bibinfo {author} {\bibfnamefont {A.}~\bibnamefont
  {Geresdi}}, \ and\ \bibinfo {author} {\bibfnamefont {M.~H.}\ \bibnamefont
  {Devoret}},\ }\href {\doibase 10.1126/science.abf0345} {\bibfield  {journal}
  {\bibinfo  {journal} {Science}\ }\textbf {\bibinfo {volume} {373}},\ \bibinfo
  {pages} {430} (\bibinfo {year} {2021})},\ \Eprint
  {http://arxiv.org/abs/https://www.science.org/doi/pdf/10.1126/science.abf0345}
  {https://www.science.org/doi/pdf/10.1126/science.abf0345} \BibitemShut
  {NoStop}%
\bibitem [{\citenamefont {Matute-Ca\~nadas}\ \emph {et~al.}(2022)\citenamefont
  {Matute-Ca\~nadas}, \citenamefont {Metzger}, \citenamefont {Park},
  \citenamefont {Tosi}, \citenamefont {Krogstrup}, \citenamefont {Nyg\aa{}rd},
  \citenamefont {Goffman}, \citenamefont {Urbina}, \citenamefont {Pothier},\
  and\ \citenamefont {Yeyati}}]{Canadas2021}%
  \BibitemOpen
  \bibfield  {author} {\bibinfo {author} {\bibfnamefont {F.~J.}\ \bibnamefont
  {Matute-Ca\~nadas}}, \bibinfo {author} {\bibfnamefont {C.}~\bibnamefont
  {Metzger}}, \bibinfo {author} {\bibfnamefont {S.}~\bibnamefont {Park}},
  \bibinfo {author} {\bibfnamefont {L.}~\bibnamefont {Tosi}}, \bibinfo {author}
  {\bibfnamefont {P.}~\bibnamefont {Krogstrup}}, \bibinfo {author}
  {\bibfnamefont {J.}~\bibnamefont {Nyg\aa{}rd}}, \bibinfo {author}
  {\bibfnamefont {M.~F.}\ \bibnamefont {Goffman}}, \bibinfo {author}
  {\bibfnamefont {C.}~\bibnamefont {Urbina}}, \bibinfo {author} {\bibfnamefont
  {H.}~\bibnamefont {Pothier}}, \ and\ \bibinfo {author} {\bibfnamefont
  {A.~L.}\ \bibnamefont {Yeyati}},\ }\href {\doibase
  10.1103/PhysRevLett.128.197702} {\bibfield  {journal} {\bibinfo  {journal}
  {Phys. Rev. Lett.}\ }\textbf {\bibinfo {volume} {128}},\ \bibinfo {pages}
  {197702} (\bibinfo {year} {2022})}\BibitemShut {NoStop}%
\bibitem [{\citenamefont {Larsen}\ \emph {et~al.}(2020)\citenamefont {Larsen},
  \citenamefont {Gershenson}, \citenamefont {Casparis}, \citenamefont
  {Kringhøj}, \citenamefont {Pearson}, \citenamefont {McNeil}, \citenamefont
  {Kuemmeth}, \citenamefont {Krogstrup}, \citenamefont {Petersson},\ and\
  \citenamefont {Marcus}}]{Larsen2020}%
  \BibitemOpen
  \bibfield  {author} {\bibinfo {author} {\bibfnamefont {T.~W.}\ \bibnamefont
  {Larsen}}, \bibinfo {author} {\bibfnamefont {M.~E.}\ \bibnamefont
  {Gershenson}}, \bibinfo {author} {\bibfnamefont {L.}~\bibnamefont
  {Casparis}}, \bibinfo {author} {\bibfnamefont {A.}~\bibnamefont {Kringhøj}},
  \bibinfo {author} {\bibfnamefont {N.~J.}\ \bibnamefont {Pearson}}, \bibinfo
  {author} {\bibfnamefont {R.~P.~G.}\ \bibnamefont {McNeil}}, \bibinfo {author}
  {\bibfnamefont {F.}~\bibnamefont {Kuemmeth}}, \bibinfo {author}
  {\bibfnamefont {P.}~\bibnamefont {Krogstrup}}, \bibinfo {author}
  {\bibfnamefont {K.~D.}\ \bibnamefont {Petersson}}, \ and\ \bibinfo {author}
  {\bibfnamefont {C.~M.}\ \bibnamefont {Marcus}},\ }\href {\doibase
  10.1103/physrevlett.125.056801} {\bibfield  {journal} {\bibinfo  {journal}
  {Physical Review Letters}\ }\textbf {\bibinfo {volume} {125}},\ \bibinfo
  {pages} {056801} (\bibinfo {year} {2020})},\ \Eprint
  {http://arxiv.org/abs/2004.03975} {2004.03975} \BibitemShut {NoStop}%
\bibitem [{\citenamefont {Sabonis}\ \emph {et~al.}(2020)\citenamefont
  {Sabonis}, \citenamefont {Erlandsson}, \citenamefont {Kringh\o{}j},
  \citenamefont {van Heck}, \citenamefont {Larsen}, \citenamefont {Petkovic},
  \citenamefont {Krogstrup}, \citenamefont {Petersson},\ and\ \citenamefont
  {Marcus}}]{SabonisLittleParks}%
  \BibitemOpen
  \bibfield  {author} {\bibinfo {author} {\bibfnamefont {D.}~\bibnamefont
  {Sabonis}}, \bibinfo {author} {\bibfnamefont {O.}~\bibnamefont {Erlandsson}},
  \bibinfo {author} {\bibfnamefont {A.}~\bibnamefont {Kringh\o{}j}}, \bibinfo
  {author} {\bibfnamefont {B.}~\bibnamefont {van Heck}}, \bibinfo {author}
  {\bibfnamefont {T.~W.}\ \bibnamefont {Larsen}}, \bibinfo {author}
  {\bibfnamefont {I.}~\bibnamefont {Petkovic}}, \bibinfo {author}
  {\bibfnamefont {P.}~\bibnamefont {Krogstrup}}, \bibinfo {author}
  {\bibfnamefont {K.~D.}\ \bibnamefont {Petersson}}, \ and\ \bibinfo {author}
  {\bibfnamefont {C.~M.}\ \bibnamefont {Marcus}},\ }\href {\doibase
  10.1103/PhysRevLett.125.156804} {\bibfield  {journal} {\bibinfo  {journal}
  {Phys. Rev. Lett.}\ }\textbf {\bibinfo {volume} {125}},\ \bibinfo {pages}
  {156804} (\bibinfo {year} {2020})}\BibitemShut {NoStop}%
\bibitem [{\citenamefont {Casparis}\ \emph {et~al.}(2019)\citenamefont
  {Casparis}, \citenamefont {Pearson}, \citenamefont {Kringh\o{}j},
  \citenamefont {Larsen}, \citenamefont {Kuemmeth}, \citenamefont {Nyg\aa{}rd},
  \citenamefont {Krogstrup}, \citenamefont {Petersson},\ and\ \citenamefont
  {Marcus}}]{Casparis2019}%
  \BibitemOpen
  \bibfield  {author} {\bibinfo {author} {\bibfnamefont {L.}~\bibnamefont
  {Casparis}}, \bibinfo {author} {\bibfnamefont {N.~J.}\ \bibnamefont
  {Pearson}}, \bibinfo {author} {\bibfnamefont {A.}~\bibnamefont
  {Kringh\o{}j}}, \bibinfo {author} {\bibfnamefont {T.~W.}\ \bibnamefont
  {Larsen}}, \bibinfo {author} {\bibfnamefont {F.}~\bibnamefont {Kuemmeth}},
  \bibinfo {author} {\bibfnamefont {J.}~\bibnamefont {Nyg\aa{}rd}}, \bibinfo
  {author} {\bibfnamefont {P.}~\bibnamefont {Krogstrup}}, \bibinfo {author}
  {\bibfnamefont {K.~D.}\ \bibnamefont {Petersson}}, \ and\ \bibinfo {author}
  {\bibfnamefont {C.~M.}\ \bibnamefont {Marcus}},\ }\href {\doibase
  10.1103/PhysRevB.99.085434} {\bibfield  {journal} {\bibinfo  {journal} {Phys.
  Rev. B}\ }\textbf {\bibinfo {volume} {99}},\ \bibinfo {pages} {085434}
  (\bibinfo {year} {2019})}\BibitemShut {NoStop}%
\bibitem [{\citenamefont {Kringhøj}\ \emph {et~al.}(2018)\citenamefont
  {Kringhøj}, \citenamefont {Casparis}, \citenamefont {Hell}, \citenamefont
  {Larsen}, \citenamefont {Kuemmeth}, \citenamefont {Leijnse}, \citenamefont
  {Flensberg}, \citenamefont {Krogstrup}, \citenamefont {Nygård},
  \citenamefont {Petersson},\ and\ \citenamefont {Marcus}}]{Kringhoj2018}%
  \BibitemOpen
  \bibfield  {author} {\bibinfo {author} {\bibfnamefont {A.}~\bibnamefont
  {Kringhøj}}, \bibinfo {author} {\bibfnamefont {L.}~\bibnamefont {Casparis}},
  \bibinfo {author} {\bibfnamefont {M.}~\bibnamefont {Hell}}, \bibinfo {author}
  {\bibfnamefont {T.~W.}\ \bibnamefont {Larsen}}, \bibinfo {author}
  {\bibfnamefont {F.}~\bibnamefont {Kuemmeth}}, \bibinfo {author}
  {\bibfnamefont {M.}~\bibnamefont {Leijnse}}, \bibinfo {author} {\bibfnamefont
  {K.}~\bibnamefont {Flensberg}}, \bibinfo {author} {\bibfnamefont
  {P.}~\bibnamefont {Krogstrup}}, \bibinfo {author} {\bibfnamefont
  {J.}~\bibnamefont {Nygård}}, \bibinfo {author} {\bibfnamefont {K.~D.}\
  \bibnamefont {Petersson}}, \ and\ \bibinfo {author} {\bibfnamefont {C.~M.}\
  \bibnamefont {Marcus}},\ }\href {\doibase 10.1103/physrevb.97.060508}
  {\bibfield  {journal} {\bibinfo  {journal} {Physical Review B}\ }\textbf
  {\bibinfo {volume} {97}},\ \bibinfo {pages} {060508} (\bibinfo {year}
  {2018})},\ \Eprint {http://arxiv.org/abs/1703.05643} {1703.05643}
  \BibitemShut {NoStop}%
\bibitem [{\citenamefont {Splitthoff}\ \emph {et~al.}(2022)\citenamefont
  {Splitthoff}, \citenamefont {Bargerbos}, \citenamefont {Grünhaupt},
  \citenamefont {Pita-Vidal}, \citenamefont {Wesdorp}, \citenamefont {Liu},
  \citenamefont {Kou}, \citenamefont {Andersen},\ and\ \citenamefont {van
  Heck}}]{splitthoff22}%
  \BibitemOpen
  \bibfield  {author} {\bibinfo {author} {\bibfnamefont {L.~J.}\ \bibnamefont
  {Splitthoff}}, \bibinfo {author} {\bibfnamefont {A.}~\bibnamefont
  {Bargerbos}}, \bibinfo {author} {\bibfnamefont {L.}~\bibnamefont
  {Grünhaupt}}, \bibinfo {author} {\bibfnamefont {M.}~\bibnamefont
  {Pita-Vidal}}, \bibinfo {author} {\bibfnamefont {J.~J.}\ \bibnamefont
  {Wesdorp}}, \bibinfo {author} {\bibfnamefont {Y.}~\bibnamefont {Liu}},
  \bibinfo {author} {\bibfnamefont {A.}~\bibnamefont {Kou}}, \bibinfo {author}
  {\bibfnamefont {C.~K.}\ \bibnamefont {Andersen}}, \ and\ \bibinfo {author}
  {\bibfnamefont {B.}~\bibnamefont {van Heck}},\ }\href@noop {} {\  (\bibinfo
  {year} {2022})},\ \Eprint {http://arxiv.org/abs/2202.08729} {arXiv:2202.08729
  [cond-mat.supr-con]} \BibitemShut {NoStop}%
\bibitem [{\citenamefont {Winkler}\ \emph {et~al.}(2019)\citenamefont
  {Winkler}, \citenamefont {Antipov}, \citenamefont {van Heck}, \citenamefont
  {Soluyanov}, \citenamefont {Glazman}, \citenamefont {Wimmer},\ and\
  \citenamefont {Lutchyn}}]{winkler2019}%
  \BibitemOpen
  \bibfield  {author} {\bibinfo {author} {\bibfnamefont {G.~W.}\ \bibnamefont
  {Winkler}}, \bibinfo {author} {\bibfnamefont {A.~E.}\ \bibnamefont
  {Antipov}}, \bibinfo {author} {\bibfnamefont {B.}~\bibnamefont {van Heck}},
  \bibinfo {author} {\bibfnamefont {A.~A.}\ \bibnamefont {Soluyanov}}, \bibinfo
  {author} {\bibfnamefont {L.~I.}\ \bibnamefont {Glazman}}, \bibinfo {author}
  {\bibfnamefont {M.}~\bibnamefont {Wimmer}}, \ and\ \bibinfo {author}
  {\bibfnamefont {R.~M.}\ \bibnamefont {Lutchyn}},\ }\href {\doibase
  10.1103/PhysRevB.99.245408} {\bibfield  {journal} {\bibinfo  {journal} {Phys.
  Rev. B}\ }\textbf {\bibinfo {volume} {99}},\ \bibinfo {pages} {245408}
  (\bibinfo {year} {2019})}\BibitemShut {NoStop}%
\bibitem [{\citenamefont {Kringh\o{}j}\ \emph
  {et~al.}(2021{\natexlab{a}})\citenamefont {Kringh\o{}j}, \citenamefont
  {Larsen}, \citenamefont {Erlandsson}, \citenamefont {Uilhoorn}, \citenamefont
  {Kroll}, \citenamefont {Hesselberg}, \citenamefont {McNeil}, \citenamefont
  {Krogstrup}, \citenamefont {Casparis}, \citenamefont {Marcus},\ and\
  \citenamefont {Petersson}}]{AndersHalfshell}%
  \BibitemOpen
  \bibfield  {author} {\bibinfo {author} {\bibfnamefont {A.}~\bibnamefont
  {Kringh\o{}j}}, \bibinfo {author} {\bibfnamefont {T.~W.}\ \bibnamefont
  {Larsen}}, \bibinfo {author} {\bibfnamefont {O.}~\bibnamefont {Erlandsson}},
  \bibinfo {author} {\bibfnamefont {W.}~\bibnamefont {Uilhoorn}}, \bibinfo
  {author} {\bibfnamefont {J.}~\bibnamefont {Kroll}}, \bibinfo {author}
  {\bibfnamefont {M.}~\bibnamefont {Hesselberg}}, \bibinfo {author}
  {\bibfnamefont {R.}~\bibnamefont {McNeil}}, \bibinfo {author} {\bibfnamefont
  {P.}~\bibnamefont {Krogstrup}}, \bibinfo {author} {\bibfnamefont
  {L.}~\bibnamefont {Casparis}}, \bibinfo {author} {\bibfnamefont
  {C.}~\bibnamefont {Marcus}}, \ and\ \bibinfo {author} {\bibfnamefont
  {K.}~\bibnamefont {Petersson}},\ }\href {\doibase
  10.1103/PhysRevApplied.15.054001} {\bibfield  {journal} {\bibinfo  {journal}
  {Phys. Rev. Applied}\ }\textbf {\bibinfo {volume} {15}},\ \bibinfo {pages}
  {054001} (\bibinfo {year} {2021}{\natexlab{a}})}\BibitemShut {NoStop}%
\bibitem [{\citenamefont {Zuo}\ \emph {et~al.}(2017)\citenamefont {Zuo},
  \citenamefont {Mourik}, \citenamefont {Szombati}, \citenamefont {Nijholt},
  \citenamefont {van Woerkom}, \citenamefont {Geresdi}, \citenamefont {Chen},
  \citenamefont {Ostroukh}, \citenamefont {Akhmerov}, \citenamefont {Plissard},
  \citenamefont {Car}, \citenamefont {Bakkers}, \citenamefont {Pikulin},
  \citenamefont {Kouwenhoven},\ and\ \citenamefont {Frolov}}]{Zou_2017}%
  \BibitemOpen
  \bibfield  {author} {\bibinfo {author} {\bibfnamefont {K.}~\bibnamefont
  {Zuo}}, \bibinfo {author} {\bibfnamefont {V.}~\bibnamefont {Mourik}},
  \bibinfo {author} {\bibfnamefont {D.~B.}\ \bibnamefont {Szombati}}, \bibinfo
  {author} {\bibfnamefont {B.}~\bibnamefont {Nijholt}}, \bibinfo {author}
  {\bibfnamefont {D.~J.}\ \bibnamefont {van Woerkom}}, \bibinfo {author}
  {\bibfnamefont {A.}~\bibnamefont {Geresdi}}, \bibinfo {author} {\bibfnamefont
  {J.}~\bibnamefont {Chen}}, \bibinfo {author} {\bibfnamefont {V.~P.}\
  \bibnamefont {Ostroukh}}, \bibinfo {author} {\bibfnamefont {A.~R.}\
  \bibnamefont {Akhmerov}}, \bibinfo {author} {\bibfnamefont {S.~R.}\
  \bibnamefont {Plissard}}, \bibinfo {author} {\bibfnamefont {D.}~\bibnamefont
  {Car}}, \bibinfo {author} {\bibfnamefont {E.~P. A.~M.}\ \bibnamefont
  {Bakkers}}, \bibinfo {author} {\bibfnamefont {D.~I.}\ \bibnamefont
  {Pikulin}}, \bibinfo {author} {\bibfnamefont {L.~P.}\ \bibnamefont
  {Kouwenhoven}}, \ and\ \bibinfo {author} {\bibfnamefont {S.~M.}\ \bibnamefont
  {Frolov}},\ }\href {\doibase 10.1103/PhysRevLett.119.187704} {\bibfield
  {journal} {\bibinfo  {journal} {Phys. Rev. Lett.}\ }\textbf {\bibinfo
  {volume} {119}},\ \bibinfo {pages} {187704} (\bibinfo {year}
  {2017})}\BibitemShut {NoStop}%
\bibitem [{\citenamefont {G{\"u}l}\ \emph {et~al.}(2014)\citenamefont
  {G{\"u}l}, \citenamefont {G{\"u}nel}, \citenamefont {L{\"u}th}, \citenamefont
  {Rieger}, \citenamefont {Wenz}, \citenamefont {Haas}, \citenamefont {Lepsa},
  \citenamefont {Panaitov}, \citenamefont {Gr{\"u}tzmacher},\ and\
  \citenamefont {Sch{\"a}pers}}]{guel2014}%
  \BibitemOpen
  \bibfield  {author} {\bibinfo {author} {\bibfnamefont {{\"O}.}~\bibnamefont
  {G{\"u}l}}, \bibinfo {author} {\bibfnamefont {H.~Y.}\ \bibnamefont
  {G{\"u}nel}}, \bibinfo {author} {\bibfnamefont {H.}~\bibnamefont {L{\"u}th}},
  \bibinfo {author} {\bibfnamefont {T.}~\bibnamefont {Rieger}}, \bibinfo
  {author} {\bibfnamefont {T.}~\bibnamefont {Wenz}}, \bibinfo {author}
  {\bibfnamefont {F.}~\bibnamefont {Haas}}, \bibinfo {author} {\bibfnamefont
  {M.}~\bibnamefont {Lepsa}}, \bibinfo {author} {\bibfnamefont
  {G.}~\bibnamefont {Panaitov}}, \bibinfo {author} {\bibfnamefont
  {D.}~\bibnamefont {Gr{\"u}tzmacher}}, \ and\ \bibinfo {author} {\bibfnamefont
  {T.}~\bibnamefont {Sch{\"a}pers}},\ }\href {\doibase 10.1021/nl502598s}
  {\bibfield  {journal} {\bibinfo  {journal} {Nano Letters}\ }\textbf {\bibinfo
  {volume} {14}},\ \bibinfo {pages} {6269} (\bibinfo {year}
  {2014})}\BibitemShut {NoStop}%
\bibitem [{\citenamefont {Beenakker}(1991)}]{Beenakkerucf}%
  \BibitemOpen
  \bibfield  {author} {\bibinfo {author} {\bibfnamefont {C.~W.~J.}\
  \bibnamefont {Beenakker}},\ }\href {\doibase 10.1103/PhysRevLett.67.3836}
  {\bibfield  {journal} {\bibinfo  {journal} {Phys. Rev. Lett.}\ }\textbf
  {\bibinfo {volume} {67}},\ \bibinfo {pages} {3836} (\bibinfo {year}
  {1991})}\BibitemShut {NoStop}%
\bibitem [{\citenamefont {Takayanagi}\ \emph {et~al.}(1995)\citenamefont
  {Takayanagi}, \citenamefont {Hansen},\ and\ \citenamefont
  {Nitta}}]{Takayanagimeso}%
  \BibitemOpen
  \bibfield  {author} {\bibinfo {author} {\bibfnamefont {H.}~\bibnamefont
  {Takayanagi}}, \bibinfo {author} {\bibfnamefont {J.~B.}\ \bibnamefont
  {Hansen}}, \ and\ \bibinfo {author} {\bibfnamefont {J.}~\bibnamefont
  {Nitta}},\ }\href {\doibase 10.1103/PhysRevLett.74.166} {\bibfield  {journal}
  {\bibinfo  {journal} {Phys. Rev. Lett.}\ }\textbf {\bibinfo {volume} {74}},\
  \bibinfo {pages} {166} (\bibinfo {year} {1995})}\BibitemShut {NoStop}%
\bibitem [{\citenamefont {Doh}\ \emph {et~al.}(2005)\citenamefont {Doh},
  \citenamefont {Dam}, \citenamefont {Roest}, \citenamefont {Bakkers},
  \citenamefont {Kouwenhoven},\ and\ \citenamefont {Franceschi}}]{dohmeso}%
  \BibitemOpen
  \bibfield  {author} {\bibinfo {author} {\bibfnamefont {Y.-J.}\ \bibnamefont
  {Doh}}, \bibinfo {author} {\bibfnamefont {J.}~\bibnamefont {Dam}}, \bibinfo
  {author} {\bibfnamefont {A.}~\bibnamefont {Roest}}, \bibinfo {author}
  {\bibfnamefont {E.}~\bibnamefont {Bakkers}}, \bibinfo {author} {\bibfnamefont
  {L.}~\bibnamefont {Kouwenhoven}}, \ and\ \bibinfo {author} {\bibfnamefont
  {S.}~\bibnamefont {Franceschi}},\ }\href {\doibase 10.1126/science.1113523}
  {\bibfield  {journal} {\bibinfo  {journal} {Science (New York, N.Y.)}\
  }\textbf {\bibinfo {volume} {309}},\ \bibinfo {pages} {272} (\bibinfo {year}
  {2005})}\BibitemShut {NoStop}%
\bibitem [{\citenamefont {Schäpers}\ \emph {et~al.}(1997)\citenamefont
  {Schäpers}, \citenamefont {Kaluza}, \citenamefont {Neurohr}, \citenamefont
  {Malindretos}, \citenamefont {Crecelius}, \citenamefont {Hart}, \citenamefont
  {Hardtdegen},\ and\ \citenamefont {Lüth}}]{shapersmeso}%
  \BibitemOpen
  \bibfield  {author} {\bibinfo {author} {\bibfnamefont {T.}~\bibnamefont
  {Schäpers}}, \bibinfo {author} {\bibfnamefont {A.}~\bibnamefont {Kaluza}},
  \bibinfo {author} {\bibfnamefont {K.}~\bibnamefont {Neurohr}}, \bibinfo
  {author} {\bibfnamefont {J.}~\bibnamefont {Malindretos}}, \bibinfo {author}
  {\bibfnamefont {G.}~\bibnamefont {Crecelius}}, \bibinfo {author}
  {\bibfnamefont {A.}~\bibnamefont {Hart}}, \bibinfo {author} {\bibfnamefont
  {H.}~\bibnamefont {Hardtdegen}}, \ and\ \bibinfo {author} {\bibfnamefont
  {H.}~\bibnamefont {Lüth}},\ }\href {\doibase 10.1063/1.120410} {\bibfield
  {journal} {\bibinfo  {journal} {Applied Physics Letters}\ }\textbf {\bibinfo
  {volume} {71}},\ \bibinfo {pages} {3575} (\bibinfo {year}
  {1997})}\BibitemShut {NoStop}%
\bibitem [{\citenamefont {Tinkham}(2004)}]{tinkham2004introduction}%
  \BibitemOpen
  \bibfield  {author} {\bibinfo {author} {\bibfnamefont {M.}~\bibnamefont
  {Tinkham}},\ }\href@noop {} {\emph {\bibinfo {title} {Introduction to
  Superconductivity}}}\ (\bibinfo  {publisher} {Courier Corporation},\ \bibinfo
  {year} {2004})\BibitemShut {NoStop}%
\bibitem [{\citenamefont {Vaitiek\ifmmode~\dot{e}\else \.{e}\fi{}nas}\ \emph
  {et~al.}(2020)\citenamefont {Vaitiek\ifmmode~\dot{e}\else \.{e}\fi{}nas},
  \citenamefont {Winkler}, \citenamefont {van Heck}, \citenamefont {Karzig},
  \citenamefont {Deng}, \citenamefont {Flensberg}, \citenamefont {Glazman},
  \citenamefont {Nayak}, \citenamefont {Krogstrup}, \citenamefont {Lutchyn},\
  and\ \citenamefont {Marcus}}]{SoleScience2020}%
  \BibitemOpen
  \bibfield  {author} {\bibinfo {author} {\bibfnamefont {S.}~\bibnamefont
  {Vaitiek\ifmmode~\dot{e}\else \.{e}\fi{}nas}}, \bibinfo {author}
  {\bibfnamefont {G.~W.}\ \bibnamefont {Winkler}}, \bibinfo {author}
  {\bibfnamefont {B.}~\bibnamefont {van Heck}}, \bibinfo {author}
  {\bibfnamefont {T.}~\bibnamefont {Karzig}}, \bibinfo {author} {\bibfnamefont
  {M.-T.}\ \bibnamefont {Deng}}, \bibinfo {author} {\bibfnamefont
  {K.}~\bibnamefont {Flensberg}}, \bibinfo {author} {\bibfnamefont {L.~I.}\
  \bibnamefont {Glazman}}, \bibinfo {author} {\bibfnamefont {C.}~\bibnamefont
  {Nayak}}, \bibinfo {author} {\bibfnamefont {P.}~\bibnamefont {Krogstrup}},
  \bibinfo {author} {\bibfnamefont {R.~M.}\ \bibnamefont {Lutchyn}}, \ and\
  \bibinfo {author} {\bibfnamefont {C.~M.}\ \bibnamefont {Marcus}},\ }\href
  {https://science.sciencemag.org/content/367/6485/eaav3392.full.pdf}
  {\bibfield  {journal} {\bibinfo  {journal} {Science}\ }\textbf {\bibinfo
  {volume} {367}} (\bibinfo {year} {2020})}\BibitemShut {NoStop}%
\bibitem [{\citenamefont {Kringh\o{}j}\ \emph
  {et~al.}(2021{\natexlab{b}})\citenamefont {Kringh\o{}j}, \citenamefont
  {Winkler}, \citenamefont {Larsen}, \citenamefont {Sabonis}, \citenamefont
  {Erlandsson}, \citenamefont {Krogstrup}, \citenamefont {van Heck},
  \citenamefont {Petersson},\ and\ \citenamefont {Marcus}}]{AndersPRL2021}%
  \BibitemOpen
  \bibfield  {author} {\bibinfo {author} {\bibfnamefont {A.}~\bibnamefont
  {Kringh\o{}j}}, \bibinfo {author} {\bibfnamefont {G.~W.}\ \bibnamefont
  {Winkler}}, \bibinfo {author} {\bibfnamefont {T.~W.}\ \bibnamefont {Larsen}},
  \bibinfo {author} {\bibfnamefont {D.}~\bibnamefont {Sabonis}}, \bibinfo
  {author} {\bibfnamefont {O.}~\bibnamefont {Erlandsson}}, \bibinfo {author}
  {\bibfnamefont {P.}~\bibnamefont {Krogstrup}}, \bibinfo {author}
  {\bibfnamefont {B.}~\bibnamefont {van Heck}}, \bibinfo {author}
  {\bibfnamefont {K.~D.}\ \bibnamefont {Petersson}}, \ and\ \bibinfo {author}
  {\bibfnamefont {C.~M.}\ \bibnamefont {Marcus}},\ }\href {\doibase
  10.1103/PhysRevLett.126.047701} {\bibfield  {journal} {\bibinfo  {journal}
  {Phys. Rev. Lett.}\ }\textbf {\bibinfo {volume} {126}},\ \bibinfo {pages}
  {047701} (\bibinfo {year} {2021}{\natexlab{b}})}\BibitemShut {NoStop}%
\bibitem [{\citenamefont {Shen}\ \emph {et~al.}(2021)\citenamefont {Shen},
  \citenamefont {Winkler}, \citenamefont {Borsoi}, \citenamefont {Heedt},
  \citenamefont {Levajac}, \citenamefont {Wang}, \citenamefont {van Driel},
  \citenamefont {Bouman}, \citenamefont {Gazibegovic}, \citenamefont
  {Op~Het~Veld}, \citenamefont {Car}, \citenamefont {Logan}, \citenamefont
  {Pendharkar}, \citenamefont {Palmstr\o{}m}, \citenamefont {Bakkers},
  \citenamefont {Kouwenhoven},\ and\ \citenamefont {van
  Heck}}]{shen2021parity}%
  \BibitemOpen
  \bibfield  {author} {\bibinfo {author} {\bibfnamefont {J.}~\bibnamefont
  {Shen}}, \bibinfo {author} {\bibfnamefont {G.~W.}\ \bibnamefont {Winkler}},
  \bibinfo {author} {\bibfnamefont {F.}~\bibnamefont {Borsoi}}, \bibinfo
  {author} {\bibfnamefont {S.}~\bibnamefont {Heedt}}, \bibinfo {author}
  {\bibfnamefont {V.}~\bibnamefont {Levajac}}, \bibinfo {author} {\bibfnamefont
  {J.-Y.}\ \bibnamefont {Wang}}, \bibinfo {author} {\bibfnamefont
  {D.}~\bibnamefont {van Driel}}, \bibinfo {author} {\bibfnamefont
  {D.}~\bibnamefont {Bouman}}, \bibinfo {author} {\bibfnamefont
  {S.}~\bibnamefont {Gazibegovic}}, \bibinfo {author} {\bibfnamefont
  {R.~L.~M.}\ \bibnamefont {Op~Het~Veld}}, \bibinfo {author} {\bibfnamefont
  {D.}~\bibnamefont {Car}}, \bibinfo {author} {\bibfnamefont {J.~A.}\
  \bibnamefont {Logan}}, \bibinfo {author} {\bibfnamefont {M.}~\bibnamefont
  {Pendharkar}}, \bibinfo {author} {\bibfnamefont {C.~J.}\ \bibnamefont
  {Palmstr\o{}m}}, \bibinfo {author} {\bibfnamefont {E.~P. A.~M.}\ \bibnamefont
  {Bakkers}}, \bibinfo {author} {\bibfnamefont {L.~P.}\ \bibnamefont
  {Kouwenhoven}}, \ and\ \bibinfo {author} {\bibfnamefont {B.}~\bibnamefont
  {van Heck}},\ }\href {\doibase 10.1103/PhysRevB.104.045422} {\bibfield
  {journal} {\bibinfo  {journal} {Phys. Rev. B}\ }\textbf {\bibinfo {volume}
  {104}},\ \bibinfo {pages} {045422} (\bibinfo {year} {2021})}\BibitemShut
  {NoStop}%
\bibitem [{\citenamefont {Groth}\ \emph {et~al.}(2014)\citenamefont {Groth},
  \citenamefont {Wimmer}, \citenamefont {Akhmerov},\ and\ \citenamefont
  {Waintal}}]{kwant}%
  \BibitemOpen
  \bibfield  {author} {\bibinfo {author} {\bibfnamefont {C.~W.}\ \bibnamefont
  {Groth}}, \bibinfo {author} {\bibfnamefont {M.}~\bibnamefont {Wimmer}},
  \bibinfo {author} {\bibfnamefont {A.~R.}\ \bibnamefont {Akhmerov}}, \ and\
  \bibinfo {author} {\bibfnamefont {X.}~\bibnamefont {Waintal}},\ }\href
  {\doibase 10.1088/1367-2630/16/6/063065} {\bibfield  {journal} {\bibinfo
  {journal} {New J. Phys.}\ }\textbf {\bibinfo {volume} {16}},\ \bibinfo
  {pages} {063065} (\bibinfo {year} {2014})},\ \bibinfo {note} {the code is
  publicly available at \url{https://kwant-project.org/}.}\BibitemShut {Stop}%
\bibitem [{\citenamefont {Ostroukh}\ \emph {et~al.}(2016)\citenamefont
  {Ostroukh}, \citenamefont {Baxevanis}, \citenamefont {Akhmerov},\ and\
  \citenamefont {Beenakker}}]{Ostroukh_2016}%
  \BibitemOpen
  \bibfield  {author} {\bibinfo {author} {\bibfnamefont {V.~P.}\ \bibnamefont
  {Ostroukh}}, \bibinfo {author} {\bibfnamefont {B.}~\bibnamefont {Baxevanis}},
  \bibinfo {author} {\bibfnamefont {A.~R.}\ \bibnamefont {Akhmerov}}, \ and\
  \bibinfo {author} {\bibfnamefont {C.~W.~J.}\ \bibnamefont {Beenakker}},\
  }\href {\doibase 10.1103/PhysRevB.94.094514} {\bibfield  {journal} {\bibinfo
  {journal} {Phys. Rev. B}\ }\textbf {\bibinfo {volume} {94}},\ \bibinfo
  {pages} {094514} (\bibinfo {year} {2016})}\BibitemShut {NoStop}%
\bibitem [{com()}]{comsol}%
  \BibitemOpen
  \href@noop {} {}\bibinfo {note} {COMSOL, Inc. [www.comsol.com]}\BibitemShut
  {NoStop}%
\bibitem [{\citenamefont {Gunel}\ \emph {et~al.}(2012)\citenamefont {Gunel},
  \citenamefont {Batov}, \citenamefont {Hardtdegen}, \citenamefont {Sladek},
  \citenamefont {Winden}, \citenamefont {Weis}, \citenamefont {Panaitov},
  \citenamefont {Gruetzmacher},\ and\ \citenamefont {Schaepers}}]{gunelmeso}%
  \BibitemOpen
  \bibfield  {author} {\bibinfo {author} {\bibfnamefont {Y.}~\bibnamefont
  {Gunel}}, \bibinfo {author} {\bibfnamefont {I.}~\bibnamefont {Batov}},
  \bibinfo {author} {\bibfnamefont {H.}~\bibnamefont {Hardtdegen}}, \bibinfo
  {author} {\bibfnamefont {K.}~\bibnamefont {Sladek}}, \bibinfo {author}
  {\bibfnamefont {A.}~\bibnamefont {Winden}}, \bibinfo {author} {\bibfnamefont
  {K.}~\bibnamefont {Weis}}, \bibinfo {author} {\bibfnamefont {G.}~\bibnamefont
  {Panaitov}}, \bibinfo {author} {\bibfnamefont {D.}~\bibnamefont
  {Gruetzmacher}}, \ and\ \bibinfo {author} {\bibfnamefont {T.}~\bibnamefont
  {Schaepers}},\ }\href {\doibase 10.1063/1.4745024} {\bibfield  {journal}
  {\bibinfo  {journal} {Journal of Applied Physics}\ }\textbf {\bibinfo
  {volume} {112}} (\bibinfo {year} {2012}),\ 10.1063/1.4745024}\BibitemShut
  {NoStop}%
\bibitem [{\citenamefont {Altshuler}\ and\ \citenamefont
  {Spivak}(1987)}]{altspivmeso}%
  \BibitemOpen
  \bibfield  {author} {\bibinfo {author} {\bibfnamefont {B.}~\bibnamefont
  {Altshuler}}\ and\ \bibinfo {author} {\bibfnamefont {B.}~\bibnamefont
  {Spivak}},\ }\href@noop {} {\bibfield  {journal} {\bibinfo  {journal} {Zh.
  Eksp. Teor. Fiz.}\ }\textbf {\bibinfo {volume} {92}} (\bibinfo {year}
  {1987})}\BibitemShut {NoStop}%
\bibitem [{\citenamefont {van Woerkom}\ \emph {et~al.}(2016)\citenamefont {van
  Woerkom}, \citenamefont {Proutski}, \citenamefont {Heck}, \citenamefont
  {Bouman}, \citenamefont {Väyrynen}, \citenamefont {Glazman}, \citenamefont
  {Krogstrup}, \citenamefont {Nygård}, \citenamefont {Kouwenhoven},\ and\
  \citenamefont {Geresdi}}]{vanWoerkom2016}%
  \BibitemOpen
  \bibfield  {author} {\bibinfo {author} {\bibfnamefont {D.}~\bibnamefont {van
  Woerkom}}, \bibinfo {author} {\bibfnamefont {A.}~\bibnamefont {Proutski}},
  \bibinfo {author} {\bibfnamefont {B.}~\bibnamefont {Heck}}, \bibinfo {author}
  {\bibfnamefont {D.}~\bibnamefont {Bouman}}, \bibinfo {author} {\bibfnamefont
  {J.}~\bibnamefont {Väyrynen}}, \bibinfo {author} {\bibfnamefont
  {L.}~\bibnamefont {Glazman}}, \bibinfo {author} {\bibfnamefont
  {P.}~\bibnamefont {Krogstrup}}, \bibinfo {author} {\bibfnamefont
  {J.}~\bibnamefont {Nygård}}, \bibinfo {author} {\bibfnamefont
  {L.}~\bibnamefont {Kouwenhoven}}, \ and\ \bibinfo {author} {\bibfnamefont
  {A.}~\bibnamefont {Geresdi}},\ }\href {\doibase 10.1038/nphys4150} {\bibfield
   {journal} {\bibinfo  {journal} {Nature Physics}\ }\textbf {\bibinfo {volume}
  {13}} (\bibinfo {year} {2016}),\ 10.1038/nphys4150}\BibitemShut {NoStop}%
\bibitem [{\citenamefont {Spanton}\ \emph {et~al.}(2017)\citenamefont
  {Spanton}, \citenamefont {Deng}, \citenamefont {Vaitiekenas}, \citenamefont
  {Krogstrup}, \citenamefont {Nygård}, \citenamefont {Marcus},\ and\
  \citenamefont {Moler}}]{Spanton2017}%
  \BibitemOpen
  \bibfield  {author} {\bibinfo {author} {\bibfnamefont {E.}~\bibnamefont
  {Spanton}}, \bibinfo {author} {\bibfnamefont {M.}~\bibnamefont {Deng}},
  \bibinfo {author} {\bibfnamefont {S.}~\bibnamefont {Vaitiekenas}}, \bibinfo
  {author} {\bibfnamefont {P.}~\bibnamefont {Krogstrup}}, \bibinfo {author}
  {\bibfnamefont {J.}~\bibnamefont {Nygård}}, \bibinfo {author} {\bibfnamefont
  {C.}~\bibnamefont {Marcus}}, \ and\ \bibinfo {author} {\bibfnamefont
  {K.}~\bibnamefont {Moler}},\ }\href {\doibase 10.1038/nphys4224} {\bibfield
  {journal} {\bibinfo  {journal} {Nature Physics}\ }\textbf {\bibinfo {volume}
  {13}} (\bibinfo {year} {2017}),\ 10.1038/nphys4224}\BibitemShut {NoStop}%
\end{thebibliography}%


%merlin.mbs apsrev4-1.bst 2010-07-25 4.21a (PWD, AO, DPC) hacked
%Control: key (0)
%Control: author (8) initials jnrlst
%Control: editor formatted (1) identically to author
%Control: production of article title (-1) disabled
%Control: page (0) single
%Control: year (1) truncated
%Control: production of eprint (0) enabled
\begin{thebibliography}{3}%
\makeatletter
\providecommand \@ifxundefined [1]{%
 \@ifx{#1\undefined}
}%
\providecommand \@ifnum [1]{%
 \ifnum #1\expandafter \@firstoftwo
 \else \expandafter \@secondoftwo
 \fi
}%
\providecommand \@ifx [1]{%
 \ifx #1\expandafter \@firstoftwo
 \else \expandafter \@secondoftwo
 \fi
}%
\providecommand \natexlab [1]{#1}%
\providecommand \enquote  [1]{``#1''}%
\providecommand \bibnamefont  [1]{#1}%
\providecommand \bibfnamefont [1]{#1}%
\providecommand \citenamefont [1]{#1}%
\providecommand \href@noop [0]{\@secondoftwo}%
\providecommand \href [0]{\begingroup \@sanitize@url \@href}%
\providecommand \@href[1]{\@@startlink{#1}\@@href}%
\providecommand \@@href[1]{\endgroup#1\@@endlink}%
\providecommand \@sanitize@url [0]{\catcode `\\12\catcode `\$12\catcode
  `\&12\catcode `\#12\catcode `\^12\catcode `\_12\catcode `\%12\relax}%
\providecommand \@@startlink[1]{}%
\providecommand \@@endlink[0]{}%
\providecommand \url  [0]{\begingroup\@sanitize@url \@url }%
\providecommand \@url [1]{\endgroup\@href {#1}{\urlprefix }}%
\providecommand \urlprefix  [0]{URL }%
\providecommand \Eprint [0]{\href }%
\providecommand \doibase [0]{http://dx.doi.org/}%
\providecommand \selectlanguage [0]{\@gobble}%
\providecommand \bibinfo  [0]{\@secondoftwo}%
\providecommand \bibfield  [0]{\@secondoftwo}%
\providecommand \translation [1]{[#1]}%
\providecommand \BibitemOpen [0]{}%
\providecommand \bibitemStop [0]{}%
\providecommand \bibitemNoStop [0]{.\EOS\space}%
\providecommand \EOS [0]{\spacefactor3000\relax}%
\providecommand \BibitemShut  [1]{\csname bibitem#1\endcsname}%
\let\auto@bib@innerbib\@empty
%</preamble>
\bibitem [{\citenamefont {Vaitiek\ifmmode~\dot{e}\else \.{e}\fi{}nas}\ \emph
  {et~al.}(2020)\citenamefont {Vaitiek\ifmmode~\dot{e}\else \.{e}\fi{}nas},
  \citenamefont {Winkler}, \citenamefont {van Heck}, \citenamefont {Karzig},
  \citenamefont {Deng}, \citenamefont {Flensberg}, \citenamefont {Glazman},
  \citenamefont {Nayak}, \citenamefont {Krogstrup}, \citenamefont {Lutchyn},\
  and\ \citenamefont {Marcus}}]{SoleScience2020}%
  \BibitemOpen
  \bibfield  {author} {\bibinfo {author} {\bibfnamefont {S.}~\bibnamefont
  {Vaitiek\ifmmode~\dot{e}\else \.{e}\fi{}nas}}, \bibinfo {author}
  {\bibfnamefont {G.~W.}\ \bibnamefont {Winkler}}, \bibinfo {author}
  {\bibfnamefont {B.}~\bibnamefont {van Heck}}, \bibinfo {author}
  {\bibfnamefont {T.}~\bibnamefont {Karzig}}, \bibinfo {author} {\bibfnamefont
  {M.-T.}\ \bibnamefont {Deng}}, \bibinfo {author} {\bibfnamefont
  {K.}~\bibnamefont {Flensberg}}, \bibinfo {author} {\bibfnamefont {L.~I.}\
  \bibnamefont {Glazman}}, \bibinfo {author} {\bibfnamefont {C.}~\bibnamefont
  {Nayak}}, \bibinfo {author} {\bibfnamefont {P.}~\bibnamefont {Krogstrup}},
  \bibinfo {author} {\bibfnamefont {R.~M.}\ \bibnamefont {Lutchyn}}, \ and\
  \bibinfo {author} {\bibfnamefont {C.~M.}\ \bibnamefont {Marcus}},\ }\href
  {https://science.sciencemag.org/content/367/6485/eaav3392.full.pdf}
  {\bibfield  {journal} {\bibinfo  {journal} {Science}\ }\textbf {\bibinfo
  {volume} {367}} (\bibinfo {year} {2020})}\BibitemShut {NoStop}%
\bibitem [{\citenamefont {Kringh\o{}j}\ \emph {et~al.}(2021)\citenamefont
  {Kringh\o{}j}, \citenamefont {Winkler}, \citenamefont {Larsen}, \citenamefont
  {Sabonis}, \citenamefont {Erlandsson}, \citenamefont {Krogstrup},
  \citenamefont {van Heck}, \citenamefont {Petersson},\ and\ \citenamefont
  {Marcus}}]{AndersPRL2021}%
  \BibitemOpen
  \bibfield  {author} {\bibinfo {author} {\bibfnamefont {A.}~\bibnamefont
  {Kringh\o{}j}}, \bibinfo {author} {\bibfnamefont {G.~W.}\ \bibnamefont
  {Winkler}}, \bibinfo {author} {\bibfnamefont {T.~W.}\ \bibnamefont {Larsen}},
  \bibinfo {author} {\bibfnamefont {D.}~\bibnamefont {Sabonis}}, \bibinfo
  {author} {\bibfnamefont {O.}~\bibnamefont {Erlandsson}}, \bibinfo {author}
  {\bibfnamefont {P.}~\bibnamefont {Krogstrup}}, \bibinfo {author}
  {\bibfnamefont {B.}~\bibnamefont {van Heck}}, \bibinfo {author}
  {\bibfnamefont {K.~D.}\ \bibnamefont {Petersson}}, \ and\ \bibinfo {author}
  {\bibfnamefont {C.~M.}\ \bibnamefont {Marcus}},\ }\href {\doibase
  10.1103/PhysRevLett.126.047701} {\bibfield  {journal} {\bibinfo  {journal}
  {Phys. Rev. Lett.}\ }\textbf {\bibinfo {volume} {126}},\ \bibinfo {pages}
  {047701} (\bibinfo {year} {2021})}\BibitemShut {NoStop}%
\bibitem [{\citenamefont {Shen}\ \emph {et~al.}(2021)\citenamefont {Shen},
  \citenamefont {Winkler}, \citenamefont {Borsoi}, \citenamefont {Heedt},
  \citenamefont {Levajac}, \citenamefont {Wang}, \citenamefont {van Driel},
  \citenamefont {Bouman}, \citenamefont {Gazibegovic}, \citenamefont {Veld},
  \citenamefont {Car}, \citenamefont {Logan}, \citenamefont {Pendharkar},
  \citenamefont {Palmstrom}, \citenamefont {Bakkers}, \citenamefont
  {Kouwenhoven},\ and\ \citenamefont {van Heck}}]{shen2021parity}%
  \BibitemOpen
  \bibfield  {author} {\bibinfo {author} {\bibfnamefont {J.}~\bibnamefont
  {Shen}}, \bibinfo {author} {\bibfnamefont {G.~W.}\ \bibnamefont {Winkler}},
  \bibinfo {author} {\bibfnamefont {F.}~\bibnamefont {Borsoi}}, \bibinfo
  {author} {\bibfnamefont {S.}~\bibnamefont {Heedt}}, \bibinfo {author}
  {\bibfnamefont {V.}~\bibnamefont {Levajac}}, \bibinfo {author} {\bibfnamefont
  {J.~Y.}\ \bibnamefont {Wang}}, \bibinfo {author} {\bibfnamefont
  {D.}~\bibnamefont {van Driel}}, \bibinfo {author} {\bibfnamefont
  {D.}~\bibnamefont {Bouman}}, \bibinfo {author} {\bibfnamefont
  {S.}~\bibnamefont {Gazibegovic}}, \bibinfo {author} {\bibfnamefont {R.~L. M.
  O.~H.}\ \bibnamefont {Veld}}, \bibinfo {author} {\bibfnamefont
  {D.}~\bibnamefont {Car}}, \bibinfo {author} {\bibfnamefont {J.~A.}\
  \bibnamefont {Logan}}, \bibinfo {author} {\bibfnamefont {M.}~\bibnamefont
  {Pendharkar}}, \bibinfo {author} {\bibfnamefont {C.~J.}\ \bibnamefont
  {Palmstrom}}, \bibinfo {author} {\bibfnamefont {E.~P. A.~M.}\ \bibnamefont
  {Bakkers}}, \bibinfo {author} {\bibfnamefont {L.~P.}\ \bibnamefont
  {Kouwenhoven}}, \ and\ \bibinfo {author} {\bibfnamefont {B.}~\bibnamefont
  {van Heck}},\ }\href@noop {} {\enquote {\bibinfo {title} {A full parity phase
  diagram of a proximitized nanowire island},}\ } (\bibinfo {year} {2021}),\
  \Eprint {http://arxiv.org/abs/2012.10118} {arXiv:2012.10118
  [cond-mat.mes-hall]} \BibitemShut {NoStop}%
\end{thebibliography}%

%%%%%%%%%% Merge with supplemental materials %%%%%%%%%%
%\widetext
%\clearpage
%\begin{center}
%\textbf{\large Supplemental Materials: Title for main text}
%\end{center}
%%%%%%%%%% Merge with supplemental materials %%%%%%%%%%
%%%%%%%%%% Prefix a "S" to all equations, figures, tables and reset the counter %%%%%%%%%%
%\setcounter{equation}{0}
%\setcounter{figure}{0}
%\setcounter{table}{0}
%\setcounter{page}{1}
%\makeatletter
%\renewcommand{\theequation}{S\arabic{equation}}
%\renewcommand{\thefigure}{S\arabic{figure}}
%\renewcommand{\bibnumfmt}[1]{[S#1]}
%\renewcommand{\citenumfont}[1]{S#1}
%%%%%%%%%% Prefix a "S" to all equations, figures, tables and reset the counter %%%%%%%%%%

\end{document}

\preprint{APS/123-QED}

\title{Supplementary Material for\\
Few-mode to mesoscopic junctions in gatemon qubits}% Force line breaks with \\

\author{Alisa~Danilenko}
\thanks{These authors contributed equally to this work}
\affiliation{Center for Quantum Devices, Niels Bohr Institute, University of Copenhagen, 2100 Copenhagen, Denmark}

\author{Deividas~Sabonis}
\thanks{These authors contributed equally to this work}

\affiliation{Center for Quantum Devices, Niels Bohr Institute, University of Copenhagen, 2100 Copenhagen, Denmark}

\author{Georg W. Winkler}
\affiliation{Microsoft Quantum, Microsoft Station Q, University of California, Santa Barbara, California 93106-6105, USA}
 
\author{Oscar~Erlandsson}
\affiliation{Center for Quantum Devices, Niels Bohr Institute, University of Copenhagen, 2100 Copenhagen, Denmark}

\author{Anders Kringhøj}
\affiliation{Center for Quantum Devices, Niels Bohr Institute, University of Copenhagen, 2100 Copenhagen, Denmark}

\author{Peter~Krogstrup}
\affiliation{Center for Quantum Devices, Niels Bohr Institute, University of Copenhagen, 2100 Copenhagen, Denmark}

\author{Charles~M.~Marcus}
\affiliation{Center for Quantum Devices, Niels Bohr Institute, University of Copenhagen, 2100 Copenhagen, Denmark}

%\date{\today}% It is always \today, today,
             %  but any date may be explicitly specified

%\begin{abstract}

%\begin{description}
%\item[Usage]
%Secondary publications and information retrieval purposes.
%\item[PACS numbers]
%May be entered using the \verb+\pacs{#1}+ command.
%\item[Structure]
%You may use the \texttt{description} environment to structure your abstract;
%use the optional argument of the \verb+\item+ command to give the category of each %item. 
%\end{description}
%\end{abstract}

%\pacs{Valid PACS appear here}% PACS, the Physics and Astronomy
                             % Classification Scheme.
%\keywords{Suggested keywords}%Use showkeys class option if keyword
                              %display desired
\maketitle

%\tableofcontents

\onecolumngrid
%%%%%%%%%% Merge with supplemental materials %%%%%%%%%%
%%%%%%%%%% Prefix a "S" to all equations, figures, tables and reset the counter %%%%%%%%%%
\setcounter{equation}{0}
\setcounter{figure}{0}
\setcounter{table}{0}
\setcounter{page}{1}
\makeatletter
\renewcommand{\theequation}{S\arabic{equation}}
\renewcommand{\thefigure}{S\arabic{figure}}
\renewcommand{\bibnumfmt}[1]{[S#1]}
\renewcommand{\citenumfont}[1]{S#1}

\subsection{\label{app:setup}Experimental setup}

\begin{figure}[h]
\includegraphics[width=6 in]{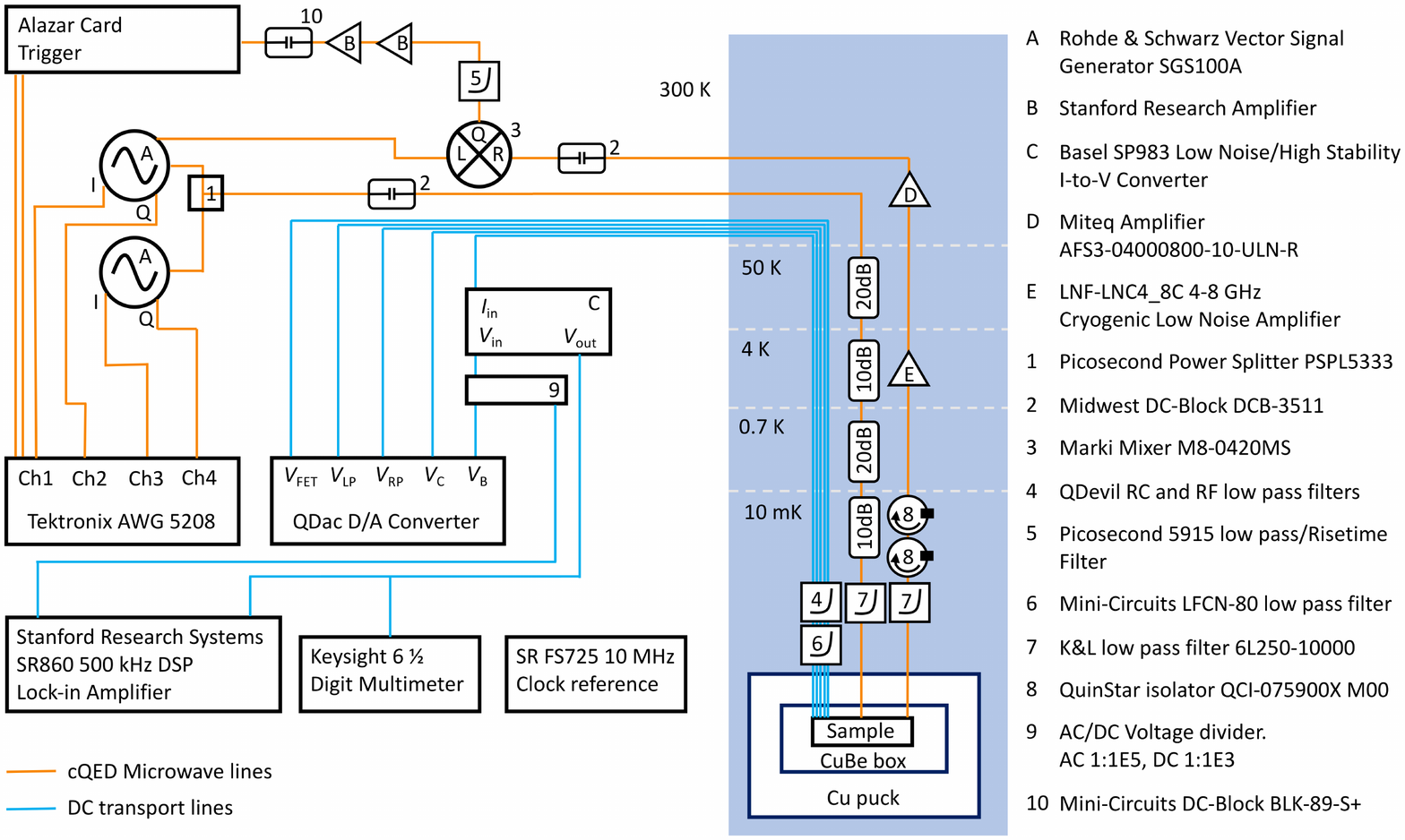}
\caption{Diagram of the experimental setup used for the experiments described in the main text. 
}
\label{fig:sup1}
\end{figure}

\begin{figure}[h]
\includegraphics[width=6 in]{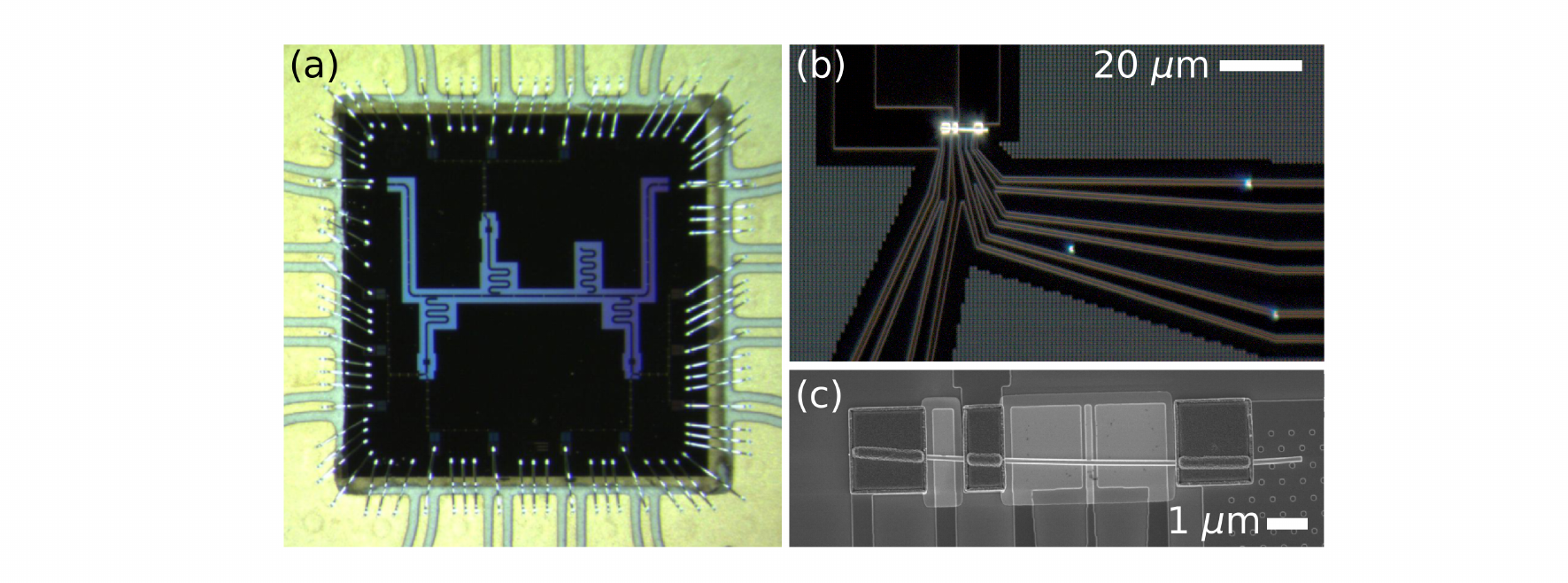}
\caption{Device 1. (a) An optical micrograph of the bonded sample, containing three qubit devices. Each device is connected to a qubit island, which is capacitively coupled to a readout resonator. Device 1, from which measurements are presented in the main text, is in the bottom right corner. (b) A magnified dark field micrograph of Device 1. The qubit island, to which the nanowire is contacted, is visible at the top of the image. (c) Scanning electron micrograph of Device 1, as seen in the main text. 
}
\label{fig:sup2}
\end{figure}

\begin{figure}[h]
\includegraphics[width=6 in]{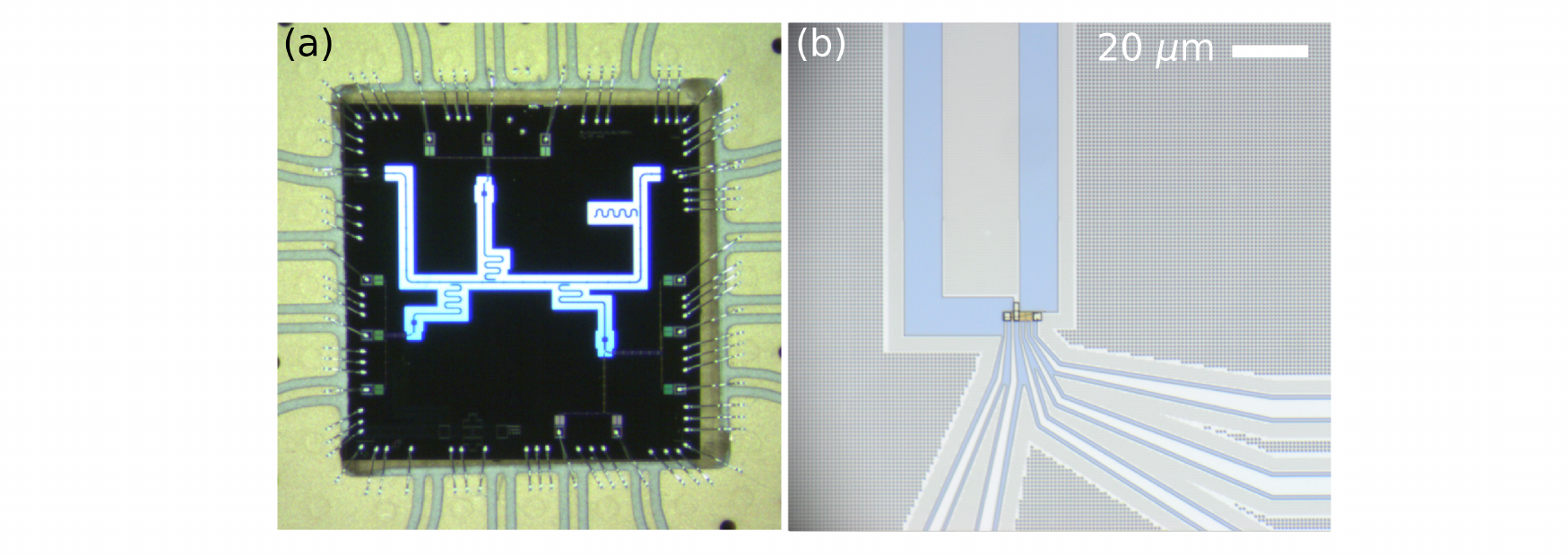}
\caption{Device 2. (a) An optical micrograph of the bonded sample, containing three qubit devices. Each device is connected to a qubit island, which is capacitively coupled to a readout resonator. Device 2, from which measurements are presented in the main text, is in the bottom right corner. (b) A magnified micrograph of Device 2. The qubit island, to which the nanowire is contacted, is visible at the top of the image, and the gate lines can be seen at the bottom.
}
\label{fig:sup3}
\end{figure}

The experimental setup used for the
measurements presented in the paper is shown in Fig.~\ref{fig:sup1}. The frequency of the readout resonator was found by transmission measurements with a vector network analyzer (VNA). The qubit measurements were then carried out using two-tone spectroscopy using a heterodyne demodulation readout circuit. Using an rf switch matrix connected to both VNA and demodulation circuit, switching was possible between the two measurement configurations. 
The experiments were carried out in a dilution refrigerator
with a base temperature of $\sim$~20~mK and a 6-1-1 T vector magnet.
Devices 1 and 2, for which data is shown in the main text, are illustrated in Figs.~\ref{fig:sup2} and \ref{fig:sup3}. An SEM image of Device 2 was not taken, to avoid detrimental effects.

\subsection{\label{app:sim}Simulation parameters}

For details of the numerical simulations see Ref.~\cite{SoleScience2020, AndersPRL2021, shen2021parity}. At the InAs/Al boundary condition we use a band offset $U_0=50\,$meV. On the bare InAs and InAs/oxide surfaces the density of interface traps is $n_\mathrm{dit}=1$e$^{12}\,$eV$^{-1}$cm$^{-2}$ with a neutral level of $\phi_\mathrm{nl}=0.1$\,eV. The magnetic field dependence of the Al gap is given by $\Delta_0(B)=\max\left(0, \Delta_0(0) (1-B^2/B_\mathrm{max}^2)\right)$ with $\Delta_0(0)=0.25$\,meV and the critical field $B_\mathrm{max}=1.4$\,T. Inside the junction region, between the Al covered segments, a random uncorrelated potential disorder is placed corresponding to a mean free path of about 50\,nm. The whole length of the simulated device is 2\,$\mu$m. The simulation does not include the spin degree of freedom, the magnetic field dependence is purely due to orbital effects.

The other simulation parameters are the same literature values as used in Ref.~\cite{SoleScience2020}: $m_\mathrm{InAs}=0.026\, m_0$, $m_\mathrm{Al}=m_0$ and $E_\mathrm{F, Al}=11.7\,$eV. Since the simulation parameters are not fine-tuned, the absolute value of qubit frequency might differ between the simulations and experiment due to differences in the electrostatic configuration, disorder or charging energy.

\subsection{\label{app:addsim}Additional simulations}

\begin{figure}[h]
\includegraphics[width=3 in]{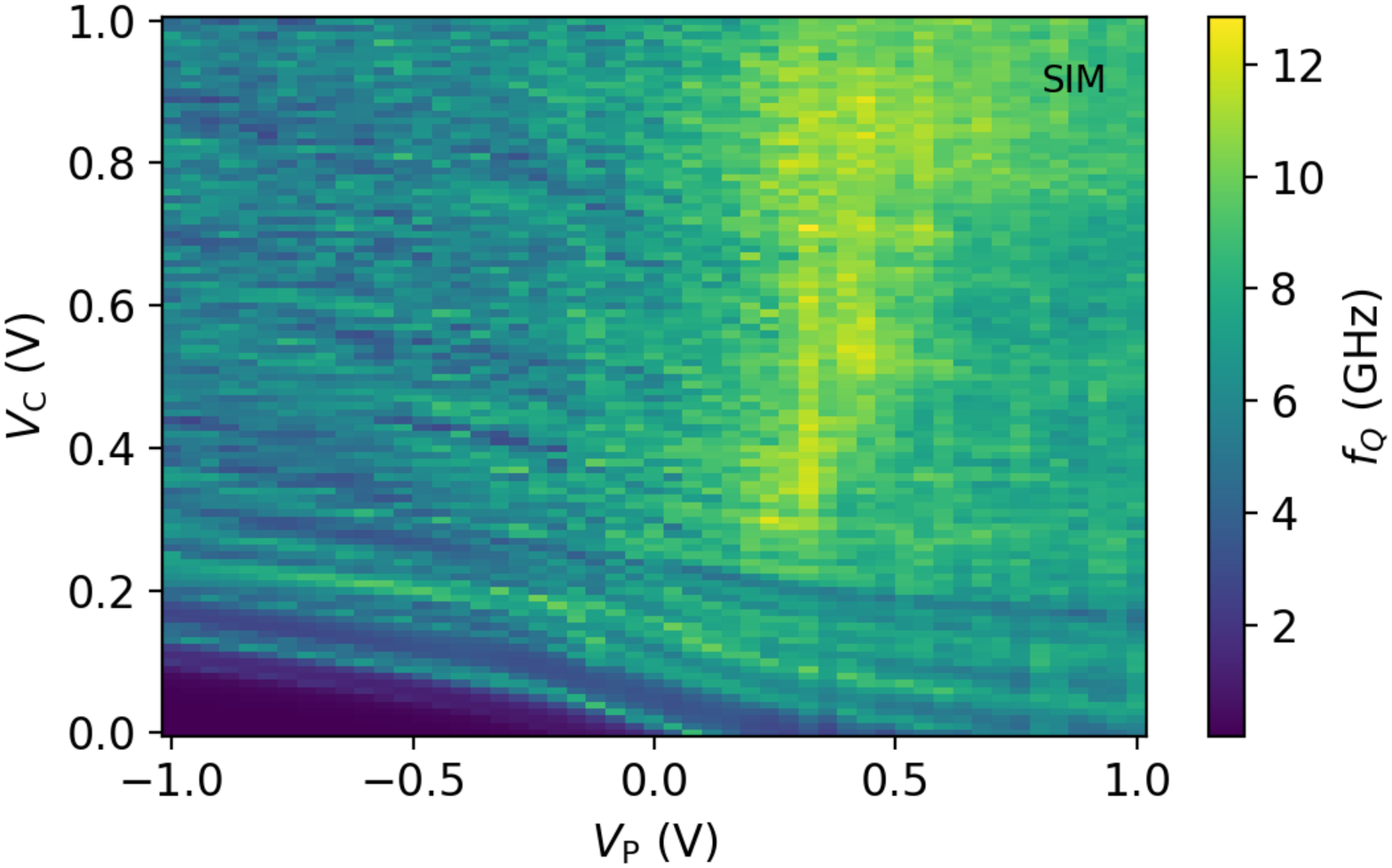}
\caption{Map of the zero field simulated qubit frequency as a function of cutter and symmetric plunger voltages ($V_\mathrm{P}=V_\mathrm{LP}=V_\mathrm{RP}$.
}
\label{fig:sup4}
\end{figure}

In Fig.~\ref{fig:sup4} we show the zero field simulated qubit frequency as a function of cutter and plunger voltages. The cutter is able to open and close the junction with relatively little cross-talk to the plungers. The behavior of the qubit frequency as a function of plunger is non-monotonic. For plungers below $V_\mathrm{P}<0.4\,\mathrm{V}$ making the plungers more positive adds more channels and the qubit frequency increases. It reaches a maximum at around $V_\mathrm{P}\sim0.4\,\mathrm{V}$. For more positive plungers the number of channels still increases, however, the induced gap collapses since the electrons are pulled away from the interface with Al. Since the added channels are only weakly proximitized, the simulated qubit frequency decreases for $V_\mathrm{P}>0.4\,\mathrm{V}$.

\begin{figure}[h]
\includegraphics[width=6 in]{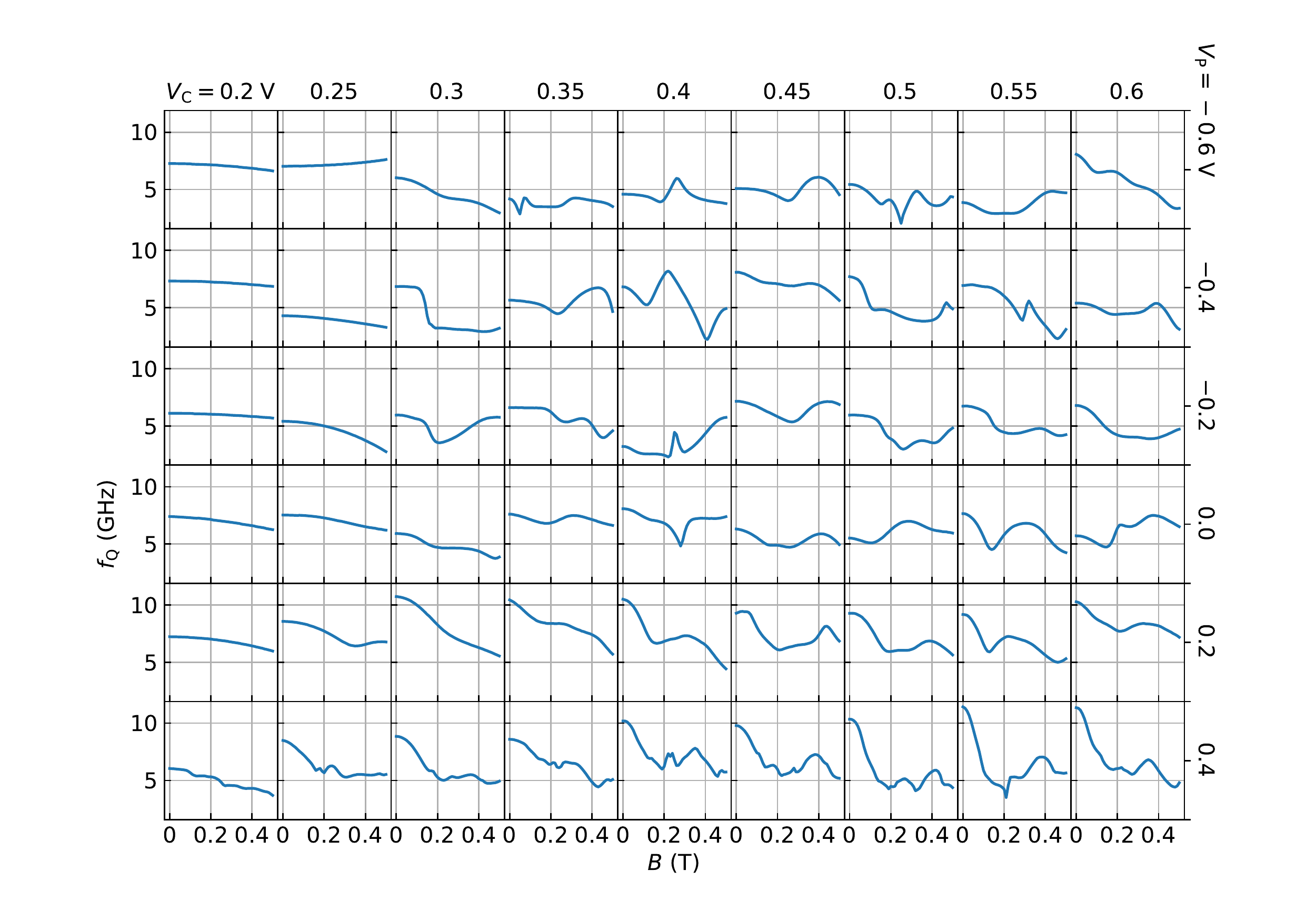}
\caption{Simulated qubit frequency as a function of magnetic field for a variety of cutter and symmetric plunger voltages ($V_\mathrm{P}=V_\mathrm{LP}=V_\mathrm{RP}$).
}
\label{fig:sup5}
\end{figure}

In Fig.~\ref{fig:sup5} we show the simulated qubit frequency as a function of magnetic field for different combinations of cutter and symmetric plunger voltages. If the junction is only barely open (i.e. $V_\mathrm{C}=0.2\,\mathrm{V}$) the qubit frequency decays monotonically with magnetic field. If the junction is more open (i.e. $V_\mathrm{C}\ge 0.3\,\mathrm{V}$) an oscillatory behavior is generally observed. Since the wire is in a multi-mode regime and different modes have different wavefunction cross-sections, the oscillations are not in every case completely regular. Furthermore, the qubit frequency never seems to crash completely before the revival but only reduces in magnitude (possibly below the limited window for which the qubit frequency is experimentally observable). 

\subsection{\label{app:setup}Additional data}

\begin{figure}[h]
\includegraphics[width=6 in]{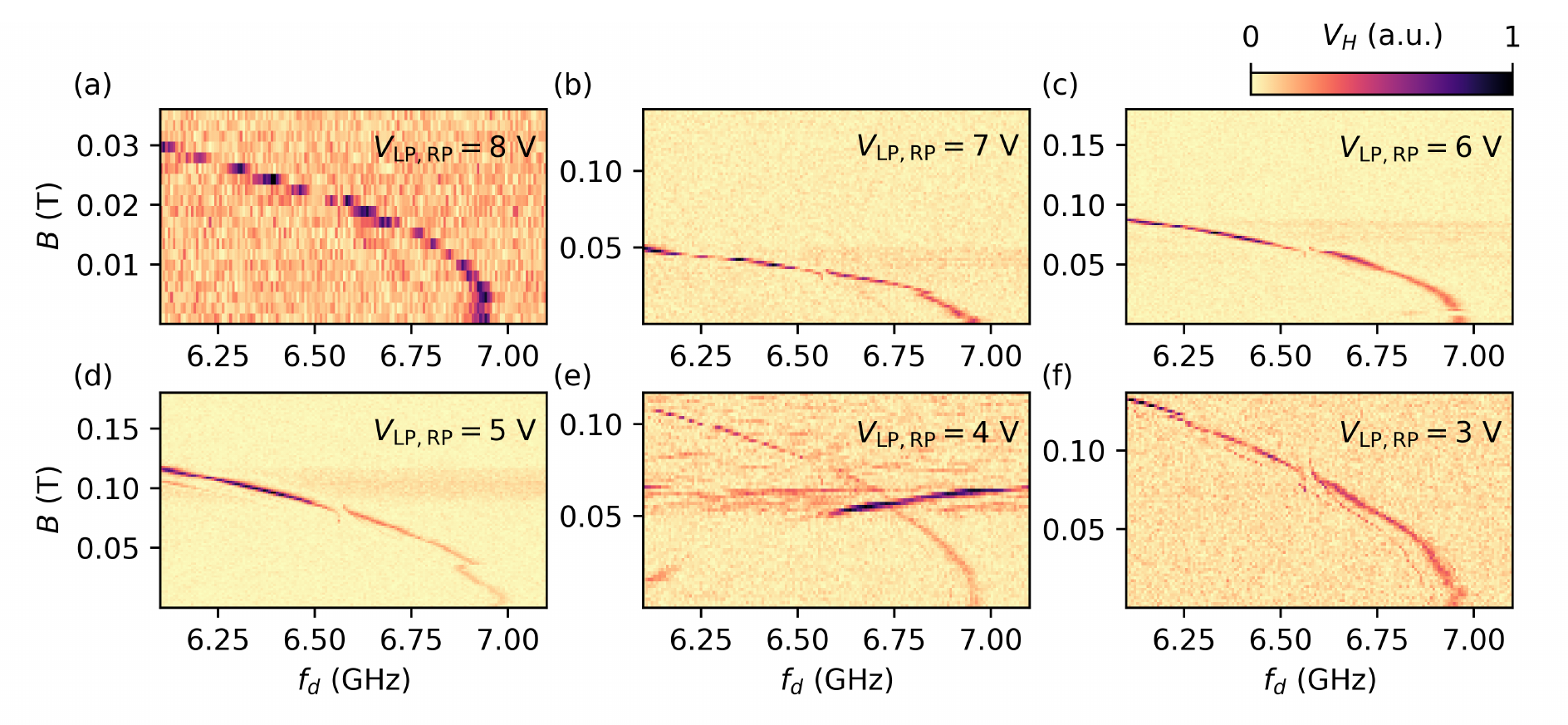}
\caption{(a) Two-tone spectroscopy of the qubit as a function of parallel magnetic
field $B$ at a range of plunger gate voltages. This data was used to extract the qubit frequencies discussed in Fig. 2 of the main text. The cutter gate was compensated slightly to keep the zero field frequency constant for all gate configurations.
}
\label{fig:sup6}
\end{figure}

\begin{figure}[h]
\includegraphics[width=3 in]{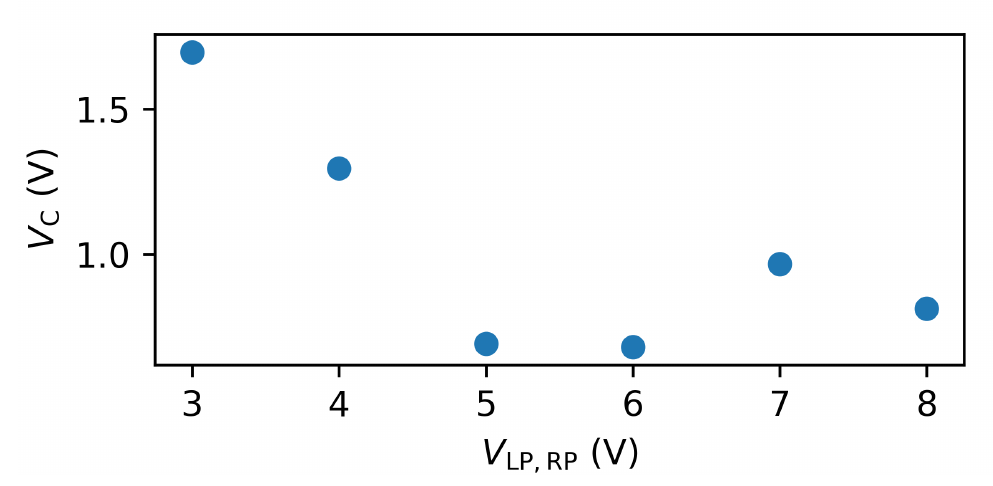}
\caption{Values of $V$\textsubscript{C} which were adjusted to compensate the zero-field frequency for the measurements of qubit frequency in parallel field, plotted as a function of the corresponding values of $V$\textsubscript{LP, RP}
}
\label{fig:sup7}
\end{figure}

In Fig. \ref{fig:sup6}, we show two-tone spectroscopy measurements as a function of parallel magnetic field $B$ at a range of $V$\textsubscript{LP, RP} voltages. An average was subtracted from each column in the measured $V$\textsubscript{H} signal to improve the qubit frequency visibility, the $Q$ and $I$ components of $V$\textsubscript{H} were added in quadrature, and finally the data was normalized. From these data, the values of $f_{\mathrm{Q}}$ shown in Fig. 2 of the main text were extracted via peak-finding. 
As mentioned in the main text, the voltage $V$\textsubscript{C} applied to the junction gate was adjusted to keep the zero field frequency close to constant as the NW gate voltages $V$\textsubscript{LP, RP} were varied. For completeness, we show the corresponding $V$\textsubscript{C} values in Fig. \ref{fig:sup7}.

\begin{figure}[h]
\includegraphics[width=6 in]{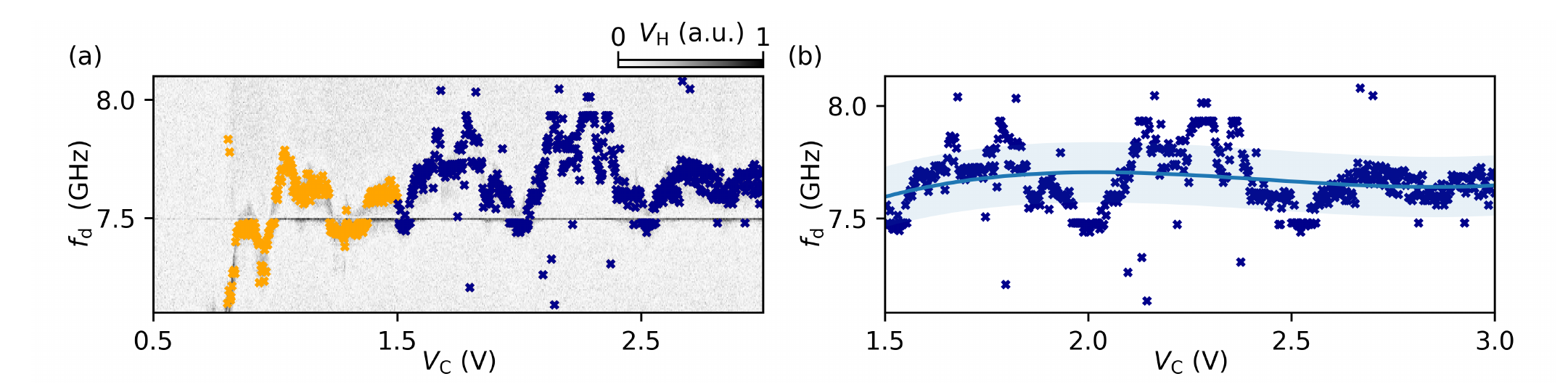}
\caption{Illustration of peak finding and fitting procedure for extraction of the mesoscopic fluctuations of the qubit frequency. (a) Shows two-tone spectroscopy as a function of drive frequency ($f_{\text{d}}$) and gate voltage $V$\textsubscript{C}, the same data as shown in Fig.~4.~(a) in the main text. Peaks extracted and used for the fitting procedure (dark blue crosses) and those extracted but unused (yellow crosses) are shown. 
(b) Shows the peaks used for the fluctuation extraction. These are fitted with a smoothed spline (solid blue line), and the standard deviation is indicated (light blue shading), covering the area within $+/- \sigma$ of the fit.
}
\label{fig:sup8}
\end{figure}

To extract the magnitude of the qubit frequency fluctuations in the many-modes section of the experiment, a two-tone spectroscopy map is taken as a function of drive frequency ($f_{\text{d}}$) and gate voltage $V$\textsubscript{C}. The $I$ and $Q$ components of the signal are measured. The mean is subtracted from each line of measurement, and the components are then added in quadrature. 
There is an accidental resonance at around 7.5 GHz, visible in Fig. \ref{fig:sup8} and also in the main text. The values of the three lines which contain the resonance are temporarily set to 0 during the peak extraction procedure, as otherwise the resonance would interfere with peak extraction. After this, the most prominent peak for every gate voltage is extracted. The extracted peak positions are shown as yellow and dark blue crosses in Fig. \ref{fig:sup8}.  We exclude from the analysis the section of the measurement where the qubit frequency starts to leave the measurement window. The peaks used for the analysis are shown as dark blue crosses. 
From this point, the analysis is identical for the experimental data shown in Fig. \ref{fig:sup8} and the simulated data sets presented in the main text.\\
The data points in the area of interest are fitted using a smoothed spline curve, with a smoothing factor 100. This number is arbitrarily selected, and does not influence the result significantly. The average frequency for the trace is then calculated by finding the mean of the data points along the smoothed interpolated trace. The fluctuation $\sigma_f$ is quantified by taking a standard deviation of the measured (or simulated) data points from the interpolated curve,
\begin{equation}
    \sigma_{f_Q} = \sqrt{{\frac{\sum(f_{\rm interp} - f_{\rm measured})^2}{N}}}
\end{equation}
where $f_{\rm interp}$ are the interpolated frequency values taken from the spline fit, $f_{\rm measured}$ are the frequency values extracted from the peak finding procedure, and $N$ is the total number of frequency data points. 

Figure~\ref{fig:sup8}~(b) shows the results of the completed fitting and extraction procedure for the experimental data. The extracted peaks from the region of interest are shown as dark blue crosses. The light blue line indicates the smoothed interpolated spline fit. The standard deviation of the data from the smoothed fit is shown as blue shading either side of the line fit.

\begin{figure}
\includegraphics[width=3 in]{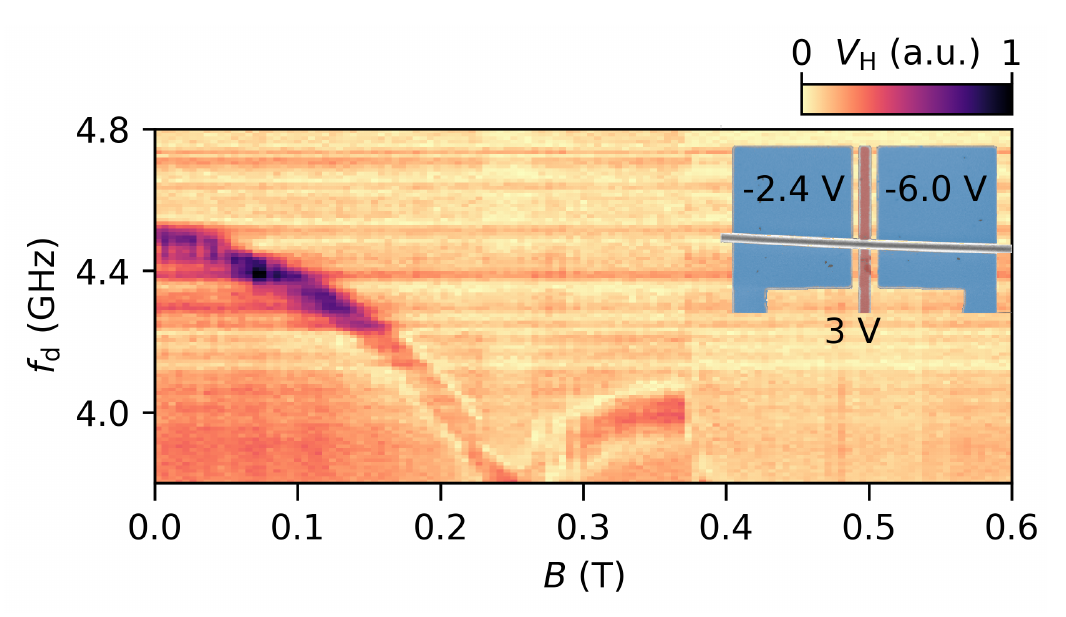}
\caption{(a) Two-tone spectroscopy
as a function of drive frequency ($f$\textsubscript{d}) and parallel magnetic field $B$, with  $V$\textsubscript{FET}~=~6~V and the other gate voltages displayed in the inset. The qubit frequency decreases monotonically at first, then at $\sim$~0.25~T a revival in the frequency is observed. There appears to be a charge switch just below 0.4~T.
}
\label{fig:sup9}
\end{figure}

In Fig.~\ref{fig:sup9} we show a similar measurement to the ones described in Fig.~3.~(a) in the main text. This data is taken on a device identical in materials and fabrication to Devices 1 and 2, and shows a non-monotonic behavior of the qubit frequency in parallel magnetic field. Unfortunately, this device was less stable than the one presented in the main text, so we observe a sudden loss of signal just below 0.4 T. It is still clear to see however that in this configuration, with a very positive junction gate, a similar modulation of the qubit frequency takes place in field. We attribute this to the same physical mechanism as described in the main text: a flux modulation in field of a few Andreev bound states in the qubit junction. 

\clearpage
\bibliography{refssup}% Produces the bibliography via BibTeX.